\documentclass[a4paper,10pt,notitlepage]{article}
\pdfoutput=1
\usepackage{datetime}
\usepackage{amsfonts,amssymb,amsmath,amsthm,hyperref,colonequals}
\usepackage[mathscr]{euscript}
\usepackage{bbm}
\usepackage[show]{ed}
\usepackage{makeidx}
\usepackage{graphicx}
\usepackage{url}
\usepackage{multirow}
\usepackage{youngtab}
\usepackage[arrow, matrix, curve]{xy}
\usepackage{bbold}
\usepackage{simplewick}
\theoremstyle{definition}
\newcommand{\beqa}{\begin{eqnarray}}
\newcommand{\eeqa}{\end{eqnarray}}
\newcommand{\beq}{\begin{equation}}
\newcommand{\eeq}{\end{equation}}

\DeclareMathOperator{\g}{\mathfrak{g}}
\DeclareMathOperator{\h}{\mathfrak{h}}
\newcommand{\osp}[2]{\ensuremath{\mathfrak{osp}\left({#1}|{#2}\right)}}

\newcommand{\OSP}[2]{\mbox{\textsc{OSP$(#1|#2)$}}}
\newcommand{\PSU}[2]{\mbox{\textsc{PSU$(#1|#2)$}}}

\newcommand{\SO}[1]{\mbox{\textsc{SO$(#1)$}}}
\newcommand{\calF}{\mathcal{F}}\newcommand{\calH}{\mathcal{H}}
\newcommand{\calS}{\mathcal{S}}\newcommand{\calP}{\mathcal{P}}

\newcommand{\calL}{\mathcal{L}}
\newcommand{\calO}{\mathcal{O}}
\newcommand{\calI}{\mathcal{I}}
\newcommand{\calN}{\mathcal{N}}
\newcommand{\calD}{\mathcal{D}}
\newcommand{\calC}{\mathcal{C}}
\newcommand{\calZ}{\mathcal{Z}}
\newcommand{\tG}{\textsf{G}}

\newcommand{\fm}{\mathfrak{m}}

\newcommand\str{\text{str}}
\newcommand\sdet{\text{sdet}}

\DeclareMathOperator{\Cas}{\mathbf{Cas}}

\DeclareMathOperator{\Ker}{\textsf{Ker}}
\DeclareMathOperator{\Img}{\textsf{Im}}
\DeclareMathOperator{\Coh}{\textsf{H}}

\newcommand{\vac}[1]{\ensuremath{\left< \, #1\, \right>}}

\newcommand{\com}[2]{\ensuremath{\left[ #1\, ,\, #2\right]}}

\newcommand{\form}[2]{\ensuremath{\left( #1, #2\right)}}

\newcommand{\floor}[1]{\ensuremath{\left\lfloor #1\right\rfloor}}

\newcommand{\tb}[1]{\textbf{#1}}

\newcommand{\G}{G}
\renewcommand{\H}{H}

\newcommand\cL{{\cal L}}
\newcommand\cP{{\cal P}}
\newcommand\cI{{\cal S}}

\renewcommand\mid{\,\vert\,}

\newcommand{\wb}[1]{\overline{#1}}

\newcommand{\defeq}{\stackrel{\textup{\tiny def}}{=}}

\DeclareFontFamily{U}{mathx}{\hyphenchar\font45}
\DeclareFontShape{U}{mathx}{m}{n}{<-> mathx10}{}
\DeclareSymbolFont{mathx}{U}{mathx}{m}{n}
\DeclareMathAccent{\widebar}{0}{mathx}{"73}

\begin{document}

\thispagestyle{empty}
\setcounter{page}{0}
\begin{flushright}\footnotesize
\texttt{DESY 13-148}\\
\texttt{HU-Mathematik-13-2013}\\
\texttt{HU-EP-13/36}\\
\vspace{0.5cm}
\end{flushright}
\setcounter{footnote}{0}

\begin{center}
{\Large\tb{\mathversion{bold}
Spectra of   Coset  Sigma Models }\par}
\vspace{15mm}

{\sc  Constantin Candu $^{a}$,
Vladimir Mitev $^{b}$,   Volker Schomerus $^{c}$ ,}\\[5mm]

{\it $^a$ Institut f\"ur Theoretische Physik,\\ Wolfgang-Pauli-Str. 27,
8093 Z\"urich, Switzerland
}\\[5mm]

{\it $^b$ Institut f\"ur Mathematik und Institut f\"ur Physik,\\ Humboldt-Universit\"at zu Berlin\\
IRIS Haus, Zum Gro{\ss}en Windkanal 6,  12489 Berlin, Germany
}\\[5mm]

{\it $^c$ DESY Hamburg, Theory Group, \\
Notkestrasse 85, D--22607 Hamburg, Germany
}\\[5mm]

\texttt{canduc@itp.phys.ethz.ch}\\
\texttt{mitev@math.hu-berlin.de}\\
\texttt{volker.schomerus@desy.de}\\[25mm]

\tb{Abstract}\\[2mm]
\end{center}
We compute the complete 1-loop spectrum of anomalous
dimensions for the bulk fields of non-linear sigma models on symmetric coset (super)spaces
$G/H$, both with and without world-sheet supersymmetry. In addition, we provide two new methods for the construction of partition functions  in the infinite radius limit
and demonstrate their efficiency in the case of (super)sphere sigma models.  Our results apply to a large number of target spaces including superspheres and superprojective spaces such as the $\mathcal{N}=2$  sigma model on $\mathbb{CP}^{3|4}$.\\


\newpage
\setcounter{page}{1}


\tableofcontents
\addtolength{\baselineskip}{5pt}

\section{Introduction}\label{sec:intro}

Non-linear $\sigma$-models (NLSM) play an important role in physics
and mathematics. While their higher dimensional versions are
non-renormalizable, NLSMs in $d >2$ dimensions are still widely
appreciated as effective field theories to study e.g.\ spontaneous
breaking of chiral symmetry in high energy physics or the low-energy
limits of numerous microscopic condensed matter models. When
$\sigma$-models are placed on a 2d world-sheet, they become
renormalizable \cite{Polyakov:1975rr,Brezin:1975sq,Friedan:1980jm}.
Initially, 2d NLSMs were
mostly studied as toy models of 4d gauge theories,
for example in order to learn about non-perturbative features and the effect
of $\theta$-terms etc., see e.g.\ \cite{D'Adda:1978uc}. But
over the last decades, numerous direct applications were
discovered.  In string theory, for example, $\sigma$-models on a
2d world-sheet are the central
ingredient of the perturbative definition, see e.g.\
\cite{Callan:1989nz} for a short review. Condensed matter
applications include spin chains, disordered metals and
superconductors, see e.g.\ \cite{Fradkin:1991nr,Tsvelik:1996zj},
while mathematicians use NLSMs in a wide range of geometric
problems, see \cite{Tseytlin:2006ak} for one example.

The properties of NLSMs depend on the choice of the
target space $\mathcal{M}$ and hence on the particular problem that is
addressed. Homogeneous target spaces are particularly relevant
for many of the aforementioned applications. In these
cases, the target (super)manifold $\mathcal{M}$ admits the transitive action of a
continuous Lie (super)group $G$. Consequently, $\mathcal{M}$ can be represented
as the coset space $\mathcal{M} =G/H$ where $H$ is the stabilizing
(super)subgroup $H \subset G$ of a point on $\mathcal{M}$. Homogeneous
(super)spaces $G/H$ for which one can find an automorphism
$\gamma:G \rightarrow G$ of order two that leaves all elements
in $H \subset G$ fixed are referred to as {\em symmetric}. According
to common folklore, NLSMs on symmetric (super)spaces tend to be
quantum integrable at least for appropriate choices of $H$, see \cite{Evans:2004mu, Babichenko:2006uc} for recent discussions and
references to older literature. Classical
integrability can be established with weaker assumptions on the
denominator subgroup, for example for supercosets $G/H$ in which the subgroup
$H$ is fixed by an automorphism of order four \cite{Bena:2003wd} (or higher \cite{Young:2005jv}).
These play an important role in the AdS/CFT correspondence and in
some cases give rise to integrable quantum theories.

In first order perturbation theory, the $\beta$-function of the
$\sigma$-model\footnote{Here the $\sigma$-model action is normalized as $\mathcal{S}=\frac{1}{2}\int d^2 x\, {\sf g}_{ij}(\phi)\partial_\mu \phi^i\partial_\mu \phi^j$.} is given by the Ricci tensor ${\sf R}$ of the
target space metric ${\sf g}$ \cite{Friedan:1980jm},
\begin{equation}
\beta_{ij} = \mu \frac{\partial {\sf g}_{ij}}{\partial \mu}
= \frac{{\sf R}_{ij}}{2\pi} + \dots
\label{eq:beta_friedan}
\end{equation}
Higher order corrections have been studied, see e.g.
\cite{Jack:1989vp} and references therein.
Many of the applications we listed above actually involve NLSMs
that are conformally invariant, i.e.\ in which the $\beta$-function
for the $\sigma$-model coupling vanishes to all orders. This is true
in particular for applications to string theory where world-sheet
conformal invariance is necessary for the consistency of the string
background. Standard examples include NLSMs on Calabi-Yau spaces.
In the context of the AdS/CFT correspondence, NLSMs on coset
superspaces such as $\PSU{2,2}{4}/\SO{1,4} \times \SO{5}$ or
$\OSP{6}{2,2}/\mbox{\textsc{U$(3)$}} \times \SO{1,3}$ have
become popular.

Conformal invariance imposes strong constraints on the target space,
in particular when the latter is  symmetric. If
the symmetry group $\G$ is simple and bosonic, it requires the
addition of a WZW term. The latter eliminates symmetry preserving
dimensionless couplings.
On the other hand, NLSMs on symmetric
superspaces $\G/\H$ with simple supergroup $G$ can be conformal, even without a WZW term.
To first order, the $\beta$-function vanishes if $G$ is one of the following supergroups
\begin{equation}
\PSU{n}{n}\ ,\quad \OSP{2n+2}{2n}\ ,\quad D(2,1;\alpha)\ .
\label{eq:special_groups}
\end{equation}
Higher order terms
impose additional restrictions on the denominator subgroup. In
absence of world-sheet supersymmetry, conformal invariance is possible only
for the following choices of $\H$ \cite{Candu:thesis, Candu:2010yg} (see also \cite{Babichenko:2006uc})
\begin{equation}\label{CNLSM}
\frac{\mathrm{OSP}(2n+2m+2|2n+2m)}{\mathrm{OSP}(2n+1|2n) \times \mathrm{OSP}(2m+1|2m)}\ , \ \   \frac{\mathrm{PSU}(n+m|n+m)}{\mathrm{S}(\mathrm{U}(n-1|n) \times \mathrm{U}(m+1|m))} \ ,
\ \
\frac{\mathrm{PSU}(2n|2n)}{ \mathrm{OSP}(2n|2n)}\ .
\end{equation}
Besides these, the  principal chiral models on the supergroups~\eqref{eq:special_groups}, which can be viewed as symmetric superspaces by the usual construction $G\simeq G\times G/ G$, are also  conformal \cite{Babichenko:2006uc, Bershadsky:1999hk, Berkovits:1999im}.
Accompanied by their different real forms, see \cite{serganova2}, the list~(\ref{eq:special_groups}, \ref{CNLSM}) exhausts all conformal $\sigma$-models on (irreducible) symmetric superspaces.
The first two families in eq.~\eqref{CNLSM} are cosets of real and complex super-Grassmannian
type. They include odd dimensional superspheres $S^{2n+1|2n}$ and complex
projective superspaces $\mathbb{CP}^{n-1|n}$ (when the parameter $m$
assumes the special value $m=0$), which are the only members in the list~(\ref{eq:special_groups}, \ref{CNLSM}) with the property of being both compact and Riemann at the same time.
The list~(\ref{eq:special_groups}, \ref{CNLSM}) can also be reproduced as
limits of GKO coset models \cite{Candu:2011hu}. World-sheet $\calN=1$
supersymmetric
extensions of the above coset models are also conformal. But once we
introduce world-sheet supersymmetry, the list (\ref{eq:special_groups}, \ref{CNLSM}) does no longer exhaust
all possibilities. To the best of our knowledge, this issue has not
been investigated in much detail. An exception are the coset models on
hermitian symmetric superspaces (see \cite{serganova2}) with numerator groups~\eqref{eq:special_groups}
\begin{equation}\label{HSS}
\frac{\mathrm{OSP}(2n+2|2n)}{\mathrm{OSP}(2n|2n) \times \mathrm{O}(2)}\ ,\quad \frac{D(2,1;\alpha)}{\mathrm{OSP}(2|2)\times \mathrm{O}(2)}\ ,\quad \frac{\mathrm{PSU}(p+q|r+s)}{\mathrm{PS}(\mathrm{U}(p|r) \times \mathrm{U}(q|s))} \ ,\quad \frac{\mathrm{OSP}(2n+2|2n)}{ \mathrm{U}(n+1|n)}
\end{equation}
and their various other real forms.
In these models the $\calN=1$ symmetry gets extended to an $\calN=2$
superconformal symmetry. Therefore, using standard non-renormalization
theorems\footnote{Usually,  for a general Calabi-Yau one must correct order by order in perturbation theory the Ricci flat metric in order to achieve conformal invariance, see \cite{Nemeschansky:1986yx}.
The particularity of the models~\eqref{HSS} is that their $\G$-invariant flat metrics do not renormalize at all, hence are known exactly to all orders in perturbation theory.}, one may argue that $\calN=1$ NLSMs on the coset spaces \eqref{HSS} are conformal to all orders, see \cite{Candu:2013??} for some indirect evidence.

In an attempt to solve conformal NLSMs, one may concentrate on the
conformal weights of fields at first.
These may be read off from the
power law decay of the 2-point functions and are expected to depend non-trivially on the (marginal) $\sigma$-model coupling.
Most of the previous systematic results on
anomalous dimensions in (not necessarily conformal) 2d NLSM on a cylinder were obtained
by Wegner \emph{et al.}\ on a case by case analysis for certain classes of symmetric spaces. Initial calculations involving operators without
derivatives \cite{Wegner:1987gu,Wegner:1987av} were performed up
to 4-loops, generalizing the pioneering 2-loop calculation of \cite{Brezin:1976an} for the $N$-vector model $\mathrm{O}(N)/\mathrm{O}(N-1)$.  The 1-loop results were later extended to fields with derivatives in \cite{Wegner:1990AD,Mall:1993wr,Wegner:1991gf}.
Anomalous dimensions of general boundary operators in conformal
NLSM on superspheres and complex projective superspaces were
studied more recently in \cite{Candu:2008vw,Candu:2008yw,
Mitev:2008yt,Candu:2009ep}. In this series of papers, all-loop
expressions were proposed for boundary conditions that preserve
the symmetries of the target space. These were tested numerically
through a relation with super-spin chains. Similar results for
periodic boundary conditions do not exist.
In this work we shall
generalize previous results on anomalous dimensions for general bulk
fields into several directions.
While our calculations are
restricted to 1-loop, they are uniformly applicable to all
symmetric (super)space models. This is achieved through a new construction of
a basis in field space and by employing the
background field expansion, see \cite{Honerkamp:1974bp,
AlvarezGaume:1981hn}.
In addition, we can incorporate $\calN=1$
world-sheet supersymmetry. Furthermore, our universal approach
also allows to obtain expressions for the full 1-loop partition function of the compact  NLSM.

Let us now outline the content of each section and the main results
of this work. In sec.~\ref{sec:sigmamodels} we shall set up most of the notations
and discuss some basics of coset (super)space $\sigma$-models, including
a construction of all their vertex operators. These vertex operators contain
two building blocks: There is a {\em zero mode} contribution formed by a
section in some homogeneous vector bundle over $\G/\H$. This is accompanied
by a {\em tail} made from products of currents and their covariant derivatives.

In sec.~\ref{sec:1loopbulk} we explain the background field expansion for symmetric
(super)spaces. The first main new result is the computation of all
1-loop anomalous dimensions in  coset $\sigma$-models on a cylinder with
periodic boundary conditions. These receive two interesting
contributions, corresponding to the two building blocks of vertex
operators we described above. The contribution associated
with zero modes is obtained from the action of the Bochner
Laplacian \cite{Pilch:1984xx} on sections of vector bundles over $\G/\H$. The scaling
behavior of the tail of currents is described by an operator with
pairwise spin-spin interactions (similar to the one appearing in the 1-loop anomalous dimensions of the perturbed WZW in \cite{Candu:2012xc}). Adding both contributions we
obtain eq.\ \eqref{eq:anom_dim_bos_bulk} for the anomalous dimension
of bulk fields at 1-loop. Though our main interest is in conformal
$\sigma$-models, we shall also briefly mention how our results extend to
cases with running $\sigma$-model coupling. In such non-conformal models,
the expression for anomalous dimensions acquires a simple additional
term, see eq.\ \eqref{eq:anom_dim_bos_bulk_nonconf}. For the $N$-vector
model our formulas reproduce the results of Wegner \cite{Wegner:1990AD}

After having computed all 1-loop anomalous dimensions for conformal
$\sigma$-models, we turn our attention to the associated partition
functions in the compact cases in sec.~\ref{sec:pf}. In a first step we need to construct the partition functions
of the free $\sigma$-models. Our general description of vertex operators in
sec.~\ref{sec:sigmamodels} lends itself to a fairly universal
construction which is also well adapted to the 1-loop deformation,
see eq.\ \eqref{eq:pf_deformed}. We explain the details of our
constructions by the example of the $N$-vector models,
show that our results agree with previously found expressions and
generalize to superspheres.

In sec.~\ref{sec:worldsheetsupersymmetry} we include $\calN=1$
world-sheet supersymmetry.
We carry out the calculation of anomalous dimensions in a manifestly supersymmetric way in complete analogy with the $\mathcal{N}=0$ case and obtain a formally identical result.
On the other hand applications include a wider class of models. While the
vanishing of the (all-loop) $\beta$-function for bosonic coset
models on $\G/\H$ imposes strong constraints on both numerator
and denominator subgroups, models with world-sheet supersymmetry
are less restrictive in their choice of the denominator
subgroup $\H$, see our discussion of the list \eqref{HSS}. The paper
concludes with an extensive list of interesting open problems
and applications.

Finally, there is a series of appendices.
Here we only mention app.~\ref{bulkspectrum}, where, as a cross-check, the partition function for the $\sigma$-models on superspheres
proposed in the main text is reproduced by an independent cohomological calculation.


\section{Coset sigma models}
\label{sec:sigmamodels}

In this section we introduce the action for $\sigma$-models on symmetric coset
superspaces $G/H$ and we describe the space of all fields. Note that certain parts of the
the following discussion can be generalized to all $G/H$ cosets, i.e.\ not only symmetric ones.

\subsection{The action}

We want to consider NLSMs on homogeneous
superspaces $\G/\H$, where the quotient is defined as the
set of right cosets of $\H$ in $\G$ through the
identification
\begin{align} g\ \sim \ gh \ \ \text{ for all } \ \  h\ \in
\ \H\ \subset\  \G\ \ .
\end{align}
Let $\g$ be the Lie superalgebra associated to $\G$. We
assume that $\g$ comes equipped with a non-degenerate
invariant bilinear form $\form{\cdot}{\cdot}$.
Similarly, let $\h$ be the Lie superalgebra associated to
$\H$. We assume that the restriction of
$\form{\cdot}{\cdot}$ to $\h$ is non-degenerate.
In this case, the orthogonal complement $\fm$ of
$\h$ in $\g$ is an $\h$-module and
one can write the following $\h$-module decomposition
$\g=\h\oplus \fm$. In
particular, this means that there are projectors $P'$ onto
$\h$ and $P=1-P'$ onto
$\fm$ which commute with the action of $\h$.

With the above requirements, the quotient $\G/\H$ can be endowed
with a $\G$-invariant metric~$\mathsf{g}$. This metric is by no
means unique and generally depends on a number of continuous
parameters.
The square root of the superdeterminant of $\mathsf{g}$
provides in the standard way a $\G$-invariant measure $\mu$ on
$\G/\H$.
With these two structures one can already
write down a purely  kinetic Lagrangian for the $\sigma$-model on
$\G/\H$ and quantize it in the path integral formalism. Inclusion
of $\theta$-terms,  WZW terms or $B$-fields requires a better
understanding of the geometry of the $\G/\H$ superspace.
We shall only consider Lagrangians with a kinetic term.
Moreover, we shall restrict to $\G/\H$ cosets which are symmetric superspaces.
For these the metric is uniquely determined up to an overall proportionality constant --- the radius squared.
Then the most general Lagrangian we shall consider can be
written in the form\footnote{The  conventions for the coordinates of an even vector are  $V=e_i V^i$, where $e_i$ is a graded basis in the tangent space.}
\begin{equation}\label{eq:lag_origin}
 \mathcal{L} = \mathsf{g}_{ij}(\phi)
 \bar \partial \phi^j \partial\phi^i
\ ,
\end{equation}
where  $\phi$ is a map from the world-sheet $\Sigma$ to the target space $\G/\H$.

There is a different way to formulate the $\sigma$-model on $\G/\H$, which makes its coset nature manifest and allows to explicitly construct the metric $\mathsf{g}$ in eq.~\eqref{eq:lag_origin}. For that purpose, let us choose a smooth (local) embedding $\imath:\G/\H\to \G$, corresponding to the choice of a coset representative $g\in \G$ for every point $g\H\in\G/\H$, and consider  instead of maps $\phi:\Sigma\to\G/\H$ the composite maps $\imath\circ \phi$. A basis set of $\g$-valued  1-forms on $\G$ which are invariant under the global left $\G$-action is provided by the
so-called Maurer-Cartan forms $\omega=g^{-1}dg$. Their pullback to the world-sheet can be decomposed as\footnote{The conventions for the coordinates of a form are $\omega = \omega_i \theta^i$, where $\theta^i$ is a graded basis in the cotangent space dual to $e_i$, i.e.\ $\theta^i(e_j)=\delta^i_j$.}
\begin{align} \label{eq:currdef}
\ \imath^{-1}(\phi(z,\bar z)) d \imath(\phi(z,\bar z)) \ &=\
\bigl[ t_{\alpha}\omega^{\alpha}_j(\phi(z,\bar z))  + t_iE^i_j(\phi(z,\bar z))\bigr] d\phi^j(z,\bar z) \nonumber\\&= J(z,\bar{z})dz+\bar J(z,\bar{z})d\bar z\ , \end{align}
where $(t_{\alpha})_{\alpha = 1}^{\dim \h}$ is a basis of
the Lie superalgebra $\h$, while $(t_{i})_{i = 1}^{\dim \fm}$ is one for the
complement $\fm$. For a symmetric superspace $G/H$
one can construct the spin connection and the vielbein
from the coefficient functions $\omega^{\alpha}_j$ and
$E^i_j$, respectively, see e.g. \cite{Kagan:2005wt}. In
terms of the currents $J$, $\bar J$ we can write
\begin{align} \label{generalLagrangian} \cL\ =\
R^2( PJ,P\bar J )
\ ,
\end{align}
where  $R$ is a proportionality coefficient defining the radius of the metric on $\G/\H$.
The relation with eq.~\eqref{eq:lag_origin} is given by
\begin{align}
\mathsf{g}_{ij}(\phi) = R^2\tG_{kl} E^k_i(\phi) E^l_j (\phi)(-1)^{|l|(|k|+|i|)}\ ,
\end{align}
where we have defined the coefficients $\tG_{kl}\defeq (t_k,t_l)$.
Any model of the form \eqref{generalLagrangian} may be considered
as a consistent $\sigma$-model on the coset space $\G/\H$. Under right
$\H$-gauge transformations $g(z,\bar z)\ \mapsto\  g(z,\bar z)h(z,\bar z)$
the currents $J(z,\bar{z})$, $\bar J(z,\bar{z})$ transform as
\begin{align} \label{rightaction}
 J\ \mapsto\
h^{-1}Jh+h^{-1}\partial
 h
\ ,\qquad
\bar J\ \mapsto\
h^{-1}\bar Jh+h^{-1}\bar \partial
 h\ . \end{align}
Since the projection $P$ on $\fm$ commutes with the
action of $\h$, the projected currents $PJ$, $P\bar J$
transform by conjugation with $h$. Hence, the Lagrangian
\eqref{generalLagrangian} is independent of how we choose
representatives in the coset space $\G/\H$, i.e.\ of the embedding $\imath$. Global left
$\G$-invariance
of the Lagrangian~\eqref{generalLagrangian} is automatic since the
Maurer-Cartan forms that we have started with are left $\G$-invariant by
construction.

\subsection{The fields}
\label{subsec:thefields}

In the metric approach~\eqref{eq:lag_origin} the space of fields is spanned by products of derivatives of vertex operators, viewed as functions in $L_2(G/H)$.
Such a representation is not very useful  for the counting of fields.
For this purpose, it is more convenient to represent the fields as products of
vertex operators with derivatives of currents, as usual in string theory.
In the approach we outlined in the second half of the previous subsection,
vertex operators are in one-to-one correspondence with square integrable sections of
homogeneous vector bundles over $\G/\H$. The relevant currents $PJ$, $P\bar J$ were used in the Lagrangian \eqref{generalLagrangian}.

In order to make this more precise, let us begin with the vertex
operators. Homogeneous vector bundles $B_\lambda\defeq G\times _H
W_\lambda$ on $\G/\H$ are associated with representations of the
denominator group $\H$ carried by a vector fiber $W_\lambda$,
where $\lambda$ is a representation label. The space of square integrable sections
may be constructed as in \cite{Bott}:
\begin{equation}
\Gamma_\lambda = \Gamma(B_\lambda,\G/\H) =
\{ \calD \in L_2(G)  : \calD(gh) = R_\lambda(h^{-1}) \calD(g)\ \forall h\in \H\}\ .
\end{equation}
Here, $R_\lambda$ denotes the representation of $\H$ on
$W_\lambda$. The space $\Gamma_\lambda$ also comes equipped with an
action of $\G$ by left multiplication, which we denote by $L$ and which commutes with the action of $\H$. Under the action of $\G$ from the left and of $H$ from the right,
the space $\Gamma_\lambda$ decomposes into the direct sum
\begin{equation} \label{eq:section}
\Gamma_\lambda \cong \bigoplus_\Lambda  n_{\lambda\Lambda}
V_\Lambda\otimes W_\lambda\ .
\end{equation}
The summation is performed over indecomposable representations $V_\Lambda$ of $\G$ labeled by $\Lambda$ and $n_{\lambda\Lambda}$ are multiplicities.

In the case of compact bosonic  Lie groups, the sum in eq.\ \eqref{eq:section}
extends over all finite dimensional irreducibles $V_\Lambda$ of $\G$ and $\Lambda$ is a dominant weight;
the multiplicities $n_{\lambda\Lambda}$ are determined by the branching of the dual representation $V^*_\Lambda=\bigoplus_\lambda n_{\lambda\Lambda}W_\lambda$ of $G$ into irreducible representations $W_\lambda$ of $\H$.
For Lie superalgebras, the decomposition  \eqref{eq:section} is more difficult to describe in general. We shall content ourselves with a formula for the character
\begin{equation}\label{eq:def_pfb}
\mathcal{Z}_{\Gamma_\lambda}(\tb{x},\tb{y}) = \str_{\Gamma_\lambda} L(\tb{x}) R_\lambda(\tb{y})
\end{equation}
of the $G\times H$ action on a compact bundle $\Gamma_\lambda$.
Here $\tb{x}$ and $\tb{y}$ are arbitrary elements of the Cartan tori of $\G$ and
$\H$, respectively. In order to calculate the counting function $\mathcal{Z}_{\Gamma_\lambda}$,
we need a
bit of background information. Under the left action of $\G$, the
space $L_2(\G)$ is known to decompose into a sum of projectives
$\cP_\Lambda$, each appearing with a multiplicity given by the dimension
of the irreducible quotient $\cI_\Lambda \subset \cP_\Lambda$. Since the restriction functor is exact, it sends projectives to projectives, i.e.\
\begin{equation}
 \cP^*_\Lambda \big\vert_{\H} = \bigoplus_\lambda n_{\lambda\Lambda} \cP_\lambda \ .
\label{eq:def_mult_sec}
\end{equation}
This decomposition formula defines the multiplicities $n_{\lambda\Lambda}$ and $\mathcal{P}_\Lambda^*$ is the dual of $\mathcal{P}_\Lambda$. Putting these facts together we obtain
the character of the bundle associated to the representation $W_\lambda=\calP_\lambda$\footnote{This formula generalizes to all submodules $W_\lambda$  of $\cP_\lambda$ if one replaces $\chi_{P_\lambda}$ by $\chi_{W_\lambda}$.
In particular, it applies to the irreducible representation $\cI_\lambda$ appearing in the socle of $\cP_\lambda$.
}
\begin{equation}\label{eq:ch_Glambda}
\mathcal{Z}_{\Gamma_\lambda}(\tb{x},\tb{y}) = \sum_\Lambda n_{\lambda\Lambda}\, \chi_{S_\Lambda}(\tb{x})  \chi_{\mathcal{P}_\lambda}(\tb{y})\ ,
\end{equation}
where $\chi_{S_\Lambda}(\tb{x})$ and $\chi_{\mathcal{P}_\lambda}(\tb{y})$ denote the
characters of the irreducible modules $S_\Lambda$ of $\G$ and the projective
$\mathcal{P}_\lambda$ of $\H$, respectively. Notice that the direct sum of all such
bundles is isomorphic to the $G\times H$ module of square integrable functions
$L_2(G)$ on the supergroup $\G$
\begin{equation}
L_2(G)=\bigoplus_\lambda \Gamma_\lambda
\label{eq:all_bd_pr}
\end{equation}
and, consequently,
\begin{equation}
\mathcal{Z}_{L_2(G)}(\tb{x},\tb{y}) = \sum_{\lambda}\mathcal{Z}_{\Gamma_\lambda}(\tb{x},\tb{y})\ ,
\label{eq:pf_all_bd_pr}
\end{equation}
where the l.h.s.\ is given by eq.\ \eqref{eq:def_pfb} after replacing $R_\lambda$ by the right action of $\H$ on $L_2(\G)$.

The vertex operators we have talked about above are associated with
sections ${\cal D}_{\Lambda\lambda} \in \Gamma_\lambda$ corresponding to the summands $V_\Lambda\otimes W_\lambda$ of eq.~\eqref{eq:section}. We shall denote them by
\begin{eqnarray}
\label{eq:vertexoperators}
\mathbb{V}_{\Lambda\lambda}(z,\bar{z})\defeq \mathbb V[{\cal D}_{\Lambda\lambda}](z,\bar z)\quad \mbox{ for } \quad  {\cal D}_{\Lambda\lambda} \in \Gamma_\lambda
\ .
\end{eqnarray}
These fields transform in the representation $W_\lambda$ under the
right action of $\H$  with the action of the group element $h$ denoted as $R_{\lambda}(h)$. In addition they carry an action of $\G$
by left translations of the sections ${\cal D}_{\Lambda\lambda}$ and transform in the representation $V_\Lambda$, with the group element $g$ acting by $L_{\Lambda}(g)$. Specifically
\beq\label{eq:sec_trpr}
\calD_{\Lambda\lambda}(gg_0)=L_{\Lambda}(g)\calD_{\Lambda\lambda}(g_0)\ , \qquad \calD_{\Lambda\lambda}(g_0h)=R_{\lambda}(h^{-1})\calD_{\Lambda\lambda}(g_0)\ ,
\eeq
for any $g$, $g_0\in \G$ and $h\in \H$.
Note that when $g_0$ coincides with the group identity $e$, consistency requires that the representation $W_\lambda$ be a submodule of the restriction of $V_\Lambda$ to $\H$ and that the two actions commute, i.e.
\beq
L_{\Lambda}(g)R_{\lambda}(h)\calD_{\Lambda\lambda}(g_0)=R_{\lambda}(h)L_{\Lambda}(g)\calD_{\Lambda\lambda}(g_0) \ .
\eeq

The second important ingredient in our construction of fields are
the currents. These were discussed quite extensively on the previous
subsection. Let us define the shorthand notations
\begin{equation}
\label{eq:defcurrents1}
\jmath \defeq  P J\ ,\qquad \bar \jmath \defeq  P \bar J\ ,\qquad A \defeq P' J\ ,\qquad \bar A \defeq P' \bar J
\end{equation}
where we recall that $P$ projects onto $\fm$ and $P'$ onto $\h$.
The gauge transformations~\eqref{rightaction} imply that $\jmath$, $\bar \jmath$ are
sections of the vector bundle associated to $\fm$.
They satisfy the equations of motion (e.o.m.) and the Maurer-Cartan equations, respectively,
\begin{equation}
\partial_A \bar{\jmath}+ \bar \partial_{\bar A} \jmath= 0 \ ,\qquad \partial_A \bar{\jmath} - \bar \partial_{\bar A} \jmath= 0\ ,
\label{eq:eom_sm}
\end{equation}
where the covariant derivative on the respective bundle is defined as
\begin{equation}
\partial_A \bar \jmath  \defeq\partial \bar \jmath+ [A,\bar \jmath]\ ,\qquad \bar \partial_{\bar A}  \jmath  \defeq\bar \partial  \jmath+ [\bar A,\jmath]\ .
\label{eq:cov_der}
\end{equation}
This is a particular case of the  covariant derivative on general sections
\begin{equation}
\partial_A \calD_{\Lambda\lambda} = [\partial +R_\lambda(A)]\calD_{\Lambda\lambda} \ ,\qquad \bar \partial_{\bar A} \calD_{\Lambda\lambda}= [\bar \partial +R_\lambda(\bar A)]\calD_{\Lambda\lambda} \ ,\qquad \calD_{\Lambda\lambda}\in \Gamma_\lambda\ .
\label{eq:cov_der_bundle}
\end{equation}

Coset fields can contain arbitrary products of higher order ``holomorphic'' covariant derivatives $\partial_A^m \jmath$ and
``anti-holomorphic'' covariant derivatives $\widebar \partial_{\bar A}^{\bar{m}} \bar\jmath$.
Mixed derivatives $\partial_A^m \widebar \partial_{\bar A}^{\bar m}$ of the currents $\jmath$, $\bar \jmath$ are not allowed due to eqs.~\eqref{eq:eom_sm}.
If  ${\mathsf{p}_{\mu}}$ denotes a projector
from $\fm^{\otimes r}$ to an indecomposable representation
$W_\mu$ of $\h$, then we can define the ``holomorphic'' composite fields
\beq
\label{eq:defproductcurrents1}
 {\jmath}_{\mu;\tb{m}}(z,\bar z)\defeq \mathsf{p}_\mu\,\jmath_{\tb m}(z,\bar{z})=\mathsf{p}_\mu\,
\bigotimes_{\rho=1}^r\partial_A^{m_i-1} \jmath(z,\bar{z})\ ,
\eeq
where $\tb{m}=\{m_1,\ldots, m_r\}$ with $m_\rho \geq m_{\rho+1}\geq 1$. 
In the same way we can define the ``anti-holomorphic'' composite fields   ${\bar\jmath}_{\bar\mu;\widebar{\tb{m}}}$ which are associated with an ordered derivative multi-index $\widebar{\tb{m}} = (\bar m_1,\dots,\bar m_{\bar{r}})$ and
projectors $\mathsf{p}_{\bar{\mu}}$ from $\fm^{\otimes \bar{r}}$ to an indecomposable representation
$W_{\bar{\mu}}$ of $\h$.
To build a basis of fields in the coset $\sigma$-model, we
must finally choose an $\h$-invariant form on the triple tensor product
$${\mathsf{c} }_{\lambda\mu\bar\mu}:\; W_\lambda\otimes W_\mu
\otimes W_{\bar\mu}  \ \rightarrow\ \mathbb{C}\, ,
$$
where, as mentioned above, $W_\lambda$ must be an indecomposable submodule of the projective representation $\calP_\lambda$.
Fields of the coset model now take the form
\beq
\label{eq:fieldscosetmodel}
 \Phi_{\boldsymbol{\Lambda}}(z,\bar z) = ({\sf d}_{\lambda\mu\bar\mu}
\mathbb V_{\Lambda\lambda}\otimes
\jmath_{\tb{m}}\otimes \bar \jmath_{\widebar{\tb{m}}})(z,\bar z) \ , \qquad \boldsymbol{\Lambda}\defeq (\Lambda, \lambda,\mu,\bar{\mu})\ ,
\eeq
where
\beq
{\sf d}_{\lambda\mu\bar\mu}\defeq {\sf c}_{\lambda\mu\bar\mu}(1\otimes \mathsf{p}_{\mu}\otimes \mathsf{p}_{\bar \mu})\ .
\eeq
By construction, these fields are invariant under the local action of the
denominator group $\H$. On the other hand, the global action of the
numerator group $\G$ is non-trivial. It is determined by the
way the sections ${\cal D}_{\Lambda\lambda}$ of eq.~\eqref{eq:vertexoperators} transform. We shall count the space of fields $\Phi$ in sec.~\ref{sec:pf}.

Before we conclude this section, let us make one more comment on
our notations. Mathematically minded readers may have wondered already
about our use of the tensor product $\otimes$ in eq.~\eqref{eq:defproductcurrents1}. In our subsequent analysis we
shall mostly work with an index free notation. To this end, the
components $\psi^\kappa$ of a field multiplet that transforms in
a representation $W_\lambda$ of $\H$ are combined into a single
object $\psi = w_\kappa \psi^\kappa$ where $w_\kappa$ is
a basis in $W_\lambda$ and summation over $\kappa$ is understood.
Now suppose we are given two such multiplets $\psi$ and $\tilde \psi$
which transform in the representations $W_\lambda$ and $W_{\tilde
\lambda}$ of the denominator subgroup $H$. Then their product $\psi
\otimes \tilde \psi = w_\kappa \otimes \tilde w_{\tilde \kappa}
\  \tilde \psi^{\tilde \kappa} \psi^\kappa $ contains all
products $ \tilde \psi^{\tilde \kappa}\psi^\kappa$ of components.
In other words, whenever we write the symbol $\otimes$ between two
field multiplets, we multiply the fields and take the tensor
product of representation spaces.

\section{Anomalous dimensions to 1-loop }
\label{sec:1loopbulk}

In this section we compute the 1-loop anomalous dimensions for $\sigma$-models on both compact and non-compact Riemann symmetric superspaces~\cite{Zirnbauer:1996zz}. We first set up the perturbative expansion in a covariant way w.r.t.\ the global $\G$ symmetry of the model using a slight modification of the background field expansion and then compute the anomalous dimensions of arbitrary
local operators. Initially these computations are performed for conformal models.
The case in which the $\sigma$-model coupling is running requires only very little additional work, so that we briefly discuss the necessary modification, even though non-conformal
models are not the main focus of our work.

\subsection{Background field expansion}
\label{sec:bfe}

For perturbative calculations it
is convenient to choose the following system of local coordinates around
a point $g_0H\in G/H$
\beq\label{eq:local_coord}
\imath:G/H\rightarrow G\ , \qquad g_0e^{i\phi}H\mapsto g_0e^{i\phi}\ ,
\eeq
where the coordinate fields $i\phi$ take values in $\fm$.
This system of coordinates has the advantage that the invariant 1-forms~\eqref{eq:currdef}
\beq\label{eq:exp_L}
\imath^{-1}d\imath=e^{-i\phi}de^{i\phi}=id\phi+\frac{1}{2}\com{\phi}{d\phi}-\frac{i}{6}\com{\phi}{\com{\phi}{d\phi}}+\cdots
\eeq
do not depend explicitly on the base point $g_0H$.
The 1-loop expansion of the
 Lagrangian~\eqref{generalLagrangian} can easily
be obtained with the help of eq.~\eqref{eq:exp_L}
\beq\label{eq:lagrangian_bfm}
\calL=R^2\form{\jmath}{\bar \jmath}=R^2\left[\form{\partial \phi}{\bar{\partial}\phi}+\frac{1}{3}\form{[\phi,\partial \phi]}{[\phi,\bar \partial \phi]}+\cdots\right]\ .
\eeq
In order to stabilize the path integral,\footnote{The Boltzmann weight is  $e^{-\mathcal{S}}$.}, we shall assume that the real form of the coset $G/H$ is chosen in such a way that
the invariant form $\form{\cdot}{\cdot}$ is negative definite when restricted to the bosonic part of $\fm$, i.e.\ $G/H$ must be a Riemannian symmetric superspace \cite{Zirnbauer:1996zz}.

We normalize the action
\beq
\label{eq:normalized1loopaction}
\calS=\frac{R^2}{2}\int\frac{d^2z}{\pi}\form{\jmath}{\bar \jmath}=\frac{R^2}{2}\int\frac{d^2z}{\pi}\left[\form{\partial \phi}{\bar{\partial}\phi}+\frac{1}{3}\form{[\phi,\partial \phi]}{[\phi,\bar \partial \phi]}+\cdots\right]
\eeq
in such a way that the free propagator takes the standard form
\beq\label{eq:prop_comp}
\vac{\phi^i(u,\bar{u})\phi^j(v,\bar{v})}_0=- (t^j,t^i)R^{-2}\log\frac{|u-v|^2}{\epsilon^2}\ ,
\eeq
where $\phi^i\defeq(\phi,t^i)$ and $(t^i,t_j)=(-1)^{|i|}(t_j,t^i)=\delta^i_j$ is the basis of $\fm$ dual to $(t_i)$.
We also have explicitly introduced a short distance cut-off $\epsilon$.
As explained in the final paragraph of the previous section we prefer to
work with elements $\phi$ rather then its components $\phi^i$.  In the
context of superalgebras this also circumvents most grading signs since
$\phi = t_i \phi^i$ is a Grassmann even combination of the graded field
components $\phi^i$. When re-written in this index free notation, the propagator~\eqref{eq:prop_comp} takes the form
\beq
\vac{\phi(u,\bar{u})\otimes \phi(v,\bar{v})}_0=-t_i\otimes t^i\, R^{-2}\log\frac{|u-v|^2}{\epsilon^2}\ .
\eeq

For the 1-loop computations, the covariant derivatives of the currents~\eqref{eq:cov_der} can be replaced with usual derivatives
$\partial$, $\bar\partial$ and it is enough to keep only the dominant terms in the currents of eq.~\eqref{eq:defcurrents1}
\beq
\label{eq:defcurrents2}
\jmath=i\partial \phi+ \dots \ ,\qquad \bar{\jmath}=i\bar{\partial}\phi+ \dots \ .
\eeq
Let us denote the \emph{dominant} terms in the composite fields \eqref{eq:defproductcurrents1} and their ``anti-holomorphic''
counterparts by
\beq
\label{eq:defproductcurrents2}
\jmath^{0}_{\tb m}\defeq \,:\bigotimes_{\rho=1}^r i\partial^{m_\rho}\phi:
\ , \qquad \bar\jmath^{0}_{\widebar{\tb{m}}}\defeq  \,:\bigotimes_{\rho=1}^{\bar r}i\bar\partial^{\bar{m}_\rho}\phi: \ ,
\eeq
where $:\cdot :$ denotes the free field normal ordering.
We also need to expand the vertex operators of eq.~\eqref{eq:vertexoperators}. This can be done as usual in the background field expansion method
\begin{align}
\mathbb V_{\Lambda\lambda}(z,\bar{z}) &= \mathcal{D}_{\Lambda\lambda}(g_0e^{it\phi(z,\bar{z})})\Big\vert_{t=1} = \sum_{n=0}^\infty \frac{1}{n!}\frac{d^n}{dt^n} \mathcal{D}_{\Lambda\lambda}(g_0e^{it\phi(z,\bar{z})})\Big\vert_{t=0}\nonumber\\
{}&=\sum_{n=0}^\infty \frac{i^n}{n!}L^n_\Lambda(\mathrm{Ad}_{g_0}\phi(z,\bar{z}))\calD_{\Lambda\lambda}(g_0)\ ,
\end{align}
where for the last equality we have used eq.~\eqref{eq:sec_trpr}.
At 1-loop we need to keep only the first two terms of the sum.
Putting things together we get the following 1-loop expansion
of a general coset field \eqref{eq:fieldscosetmodel} around an
arbitrary point $g_0H$
\beq
\label{eq:1loopfields}
\Phi_{\boldsymbol{\Lambda}}(z,\bar{z}\mid g_0)=\, :{\sf d}_{\lambda\mu\nu}\left[\calD_{\Lambda\lambda}(g_0)+iL_\Lambda(\mathrm{Ad}_{g_0}\phi(z,\bar{z}))\calD_{\Lambda\lambda}(g_0)\right]\otimes \jmath^{0}_{\tb m}(z)\otimes
\bar \jmath^{0}_{\widebar{\tb m}}(\bar z):\, + \dots \ .
\eeq
Our formulas \eqref{eq:normalized1loopaction} and \eqref{eq:1loopfields} for the
1-loop approximation of the Lagrangian and of the fields in the $\sigma$-model provide
the basic input for all our perturbative computations below.

The quantity we are most interested in is the 1-loop correction to the
anomalous dimension of coset fields. This is encoded in the 2-point
functions,
\begin{multline}
\label{eq:defFbulk}
\vac{\Phi_{\boldsymbol{\Lambda}}(u,\bar{u})\otimes \Phi_{\boldsymbol{\Xi}}(v,\bar{v})}
=\int_{G/H}d\mu(g_0H)\vac{\Phi_{\boldsymbol{\Lambda}}(u,\bar{u}\mid g_0)\otimes \Phi_{\boldsymbol{\Xi}}(v,\bar{v}\mid g_0)e^{-\calS_{\text{int}}}}_{0,c}\ ,
\end{multline}
where the subscript $c$ indicates the removal of vacuum bubbles. In
computing the 1-loop correction to the 2-point function we can use the
following expression for the interaction, see eq.\
\eqref{eq:normalized1loopaction},
\beq
\calS_{\text{int}}=\int \frac{d^2z}{\pi}\Omega(z,\bar{z})\ ,\qquad \Omega(z,\bar{z})\defeq \frac{R^2}{6}: \form{[\phi,\partial \phi]}{[\phi,\bar \partial \phi]}:\ .
\eeq
If we expand the exponential and use the expression \eqref{eq:1loopfields}
to separate the fields into a background and a quantum piece, the tree-level
contribution to the 2-point functions takes the form,
\begin{equation}
\label{eq:0loopcorrelationfunction}
\vac{\Phi_{\boldsymbol{\Lambda}}(u,\bar{u})\otimes \Phi_{\boldsymbol{\Xi}}(v,\bar{v})}_0=\int_{G/H}d\mu(g_0H)\,({\sf d}_{\lambda\mu\bar\mu}\otimes {\sf d}_{\xi\eta\bar{\eta}})I_0\ ,
\end{equation}
where $\boldsymbol{\Xi}=(\Xi,\xi,\eta,\bar \eta)$ and
\begin{equation}
I_0\defeq\vac{\calD_{\Lambda\lambda}(g_0)\otimes \jmath^{0}_{\tb m} (u)\otimes \wb\jmath^{0}_{\widebar{\tb m}} (\bar{u})\otimes
\calD_{\Xi\xi}(g_0)\otimes
\jmath^{0}_{\tb{n}}(v)\otimes \bar \jmath^{0}_{\widebar{\tb{n}}}(\bar{v})}_0\ ,
\end{equation}
while the 1-loop correction is given by
\beq
\label{eq:1loopcorrelationfunction}
\vac{\Phi_{\boldsymbol{\Lambda}}(u,\bar{u})\otimes \Phi_{\boldsymbol{\Xi}}(v,\bar{v})}_1=\int_{G/H}d\mu(g_0H)\,({\sf d}_{\lambda\mu\bar\mu}\otimes {\sf d}_{\xi\eta\bar{\eta}})\left(I'_1+I''_1\right)
\eeq
where
\begin{align}
I'_1&\defeq -\int_{\mathbb{C}_{\epsilon}}\frac{d^2z}{\pi}\,\vac{\calD_{\Lambda\lambda}(g_0)\otimes
\jmath^0_{\tb{m}}(u)\otimes
\bar\jmath^0_{\widebar{\tb{m}}}(\bar{u})\otimes \calD_{\Xi\xi}(g_0)\otimes
\jmath^0_{\tb{n}}(v)\otimes \bar\jmath^0_{\widebar{\tb{n}}}(\bar{v})\,\Omega(z,\bar{z})}_0\ ,\nonumber\\
I''_1 &\defeq-\Big\langle\, :L_\Lambda(\mathrm{Ad}_{g_0}\phi(u,\bar{u}))\calD_{\Lambda\lambda}(g_0)\otimes
\jmath^0_{\tb{m}}(u)\otimes
\bar\jmath^0_{\widebar{\tb{m}}}(\bar{u}):\otimes \label{eq:I1I2}
\\{}&\hspace{5.6cm}\otimes :L_{\Xi}(\mathrm{Ad}_{g_0}\phi(v,\bar{v}))\calD_{\Xi\xi}(g_0)\,\otimes
\jmath^0_{\tb{n}}(v)\otimes \bar\jmath^0_{\widebar{\tb{n}}}(\bar{v}):\,\Big\rangle \ .
\notag
\end{align}
The integral in $I'_1$ is regularized by a short distance cut-off $\epsilon$
\begin{equation}
\mathbb{C}_\epsilon\defeq\{\,z\in \mathbb C\ \mid\ \epsilon\leq|z-u|,\, \epsilon\leq |z-v|\,\}\ .
\end{equation}
In order to extract the 1-loop anomalous dimensions, we need to compute only those pieces of eq.\ \eqref{eq:1loopcorrelationfunction} that contain logarithms of the UV cut-off $\epsilon$. All other terms would matter if we were attempting to construct the
eigenvectors of the 1-loop dilatation operator. For the eigenvalues, however, they
are not relevant.

\subsection{The 1-loop dilatation operator}
\label{subsec:1loopdilatation}

The calculation of anomalous dimensions depends somewhat on whether the $\sigma$-model is conformal or not. In the conformal case,
the kinetic term $(\jmath,\bar{\jmath})$ is an operator of dimension
$(1,1)$ in the interacting theory and the radius $R$ is a dimensionless
parameter of the model. This will be the case if and only if the
$\h$-Casimir of $\fm$ together with all higher order Casimirs vanishes, see \cite{Candu:thesis,Candu:2010yg}.

The anomalous dimensions $\delta \mathbf{h}$ can be read off from the
following expression for the 1-loop correction to the 2-point function,
\begin{equation}
\vac{\Phi_{\boldsymbol{\Lambda}}(u,\bar{u})\otimes \Phi_{\boldsymbol{\Xi}}(v,\bar{v})}_1 = \vac{2\delta\mathbf{h}\cdot \Phi_{\boldsymbol{\Lambda}}(u,\bar{u})\otimes \Phi_{\boldsymbol{\Xi}}(v,\bar{v})}_0\, \log\left|\frac{\epsilon}{u-v}\right|^2 + \dots
\label{eq:anom_dim_conf}
\end{equation}
where $\delta\mathbf{h}$ is a $\G$-invariant operator acting on the
representation indices of $\Phi_{\boldsymbol{\Lambda}}$. To remove the cut-off $\epsilon$ in
favor of an arbitrary  scale $\mu$ one must renormalize the fields as $\Phi =
Z_\Phi \cdot \Phi^{\text{ren}}$ with
\begin{equation}
 Z_\Phi = 1 + 2\delta\mathbf{ h}_\Phi \log {\mu} \epsilon+\mathcal{O}(R^{-4})\ .
\label{eq:anom_dim_nonconf}
\end{equation}
In the non-conformal case, the kinetic term $(\jmath,\bar\jmath)$, as it stands,
fails to be marginal in the interacting theory, i.e.\ it acquires an anomalous dimension. The perturbative calculations will be valid at high/low energies if
it is marginally relevant/irrelevant. To make the operator $(\jmath,\bar\jmath)$  marginal at 1-loop one must also renormalize the coordinate fields\footnote{Note that while $\phi$ is a building block of the coset model, it is not by itself an observable of the theory.} $\phi\to Z\phi$.
More generally, composite operators must be  renormalized according to the formula
\begin{equation}
Z_\Phi \cdot \Phi^{\mathrm{ren}} \defeq  \Phi\big\vert_{\phi\to Z\phi}\ ,
\label{eq:wave_func_renorm}
\end{equation}
in such a way that all the dependence of the 2-point functions on the UV cut-off $\epsilon$ is eliminated in favor of a renormalization scale $ {\mu}$. The 1-loop
anomalous dimensions are then  again given by eq.~\eqref{eq:anom_dim_nonconf},
where now the main difference is that the renormalized coupling $R$ has a non-trivial $\beta$-function. The latter is computed by requiring that the bare radius  $R Z^{-1}$ is independent of the  renormalization scale $\mu$
\begin{equation}
{\mu}\frac{d}{d{\mu}} R^2Z^{-2} = 0\ .
\label{eq:def_beta}
\end{equation}
In the following we shall first address the case in which the radius $R$ does
not run. The effect of a running $\sigma$-model coupling $R^{-2}$, which is relevant
for all coset $\sigma$-models in which the numerator group $G$ has non-vanishing
dual Coxeter number, can be easily incorporated. We shall discuss the small
modifications in a second subsection.

\subsubsection{Conformal case}

As we discussed at the end of sec.~\ref{sec:bfe}, the 1-loop correction \eqref{eq:1loopcorrelationfunction} to the 2-point function consists of
two separate contributions, namely $I'_1$ and $I''_1$. The first term $I'_1$ involves a single
insertion of the interaction $\Omega$.
Before we begin to evaluate this term, let us point out that the tree-level
piece $I_0$ in eq.\ \eqref{eq:0loopcorrelationfunction} is non-zero if and
only if the vectors $\tb{m}$ and $\tb{n}$ have the same number of
components, namely $r$. The same must hold for $\widebar{\tb{m}}$ and $\widebar{\tb{n}}$.
We shall denote the corresponding number of components by $\bar{r}$.
Moreover, $I_1'$ vanishes unless both $r$ and $\bar r$ are strictly bigger
than zero. Under these restrictions, we find
\begin{align}\notag
&\int_{\mathbb{C}_{\epsilon}}d^2z\,\vac{
\jmath^0_{\tb{m}}(u)\otimes
\bar\jmath^0_{\widebar{\tb{m}}}(\bar{u})\otimes
\jmath^0_{\tb{n}}(v)\otimes \bar\jmath^0_{\widebar{\tb{n}}}(\bar{v})\,\Omega(z,\bar{z})}_0=\Pi\cdot\Bigg[
\sum_{\rho,\sigma=1}^r\sum_{\bar\rho,\bar\sigma=1}^{\bar r}
\Big\langle\,\jmath^0_{\tb{m}_\rho}(u)\otimes
\bar\jmath^0_{\widebar{\tb{m}}_{\bar \rho}}(\bar{u})\,\otimes\\
&\otimes\, \jmath^0_{\tb{n}_\sigma}(v)\otimes \bar\jmath^0_{\widebar{\tb{n}}_{\bar\sigma}}(\bar{v})\,\Big\rangle_0
\otimes
\int_{\mathbb{C}_\epsilon}\frac{d^2z}{\pi}
\vac{
\partial^{m_\rho}\phi(u)\otimes
\bar\partial^{\bar{m}_{\bar\rho}}\phi(\bar{u})\otimes
\partial^{n_{\sigma}}\phi(v)\otimes\bar\partial^{\bar{n}_{\bar\sigma}}\phi(\bar{v})\,\Omega(z,\bar{z})}_0 \Bigg]\ .
\label{eq:sumdecomposition}
\end{align}
Here, $\jmath^0_{\tb{m}_\rho}$ denotes the tensor product \eqref{eq:defproductcurrents2} with the $\rho$-th factor removed and we introduced a permutation $\Pi$ that acts on
a tensor power of $\mathfrak{m}$. Its purpose is to bring the four factors $\mathfrak{m}$ that are associated with the four copies of $\phi$ under the integral back into the original
positions $\rho,\sigma$ and $\bar\rho,\bar\sigma$.

If it were not for the insertion of the interaction term $\Omega$, the four
field correlation function in our expression for $I'_1$ would be very easy to
evaluate. The answer is given by
\begin{multline}
\label{eq:basiccorrelator1loop}
\vac{\partial^{m}\phi(u)\otimes  \bar{\partial}^{\bar m}\phi(\bar{u})\otimes\partial^{n}\phi(v)\otimes\bar{\partial}^{\bar n}\phi(\bar{v})}_0=\\=
(-1)^{|i||j|}t_i \otimes t_j\otimes t^i\otimes t^j\,\frac{(-1)^{m+\bar m}(m+n-1)!(\bar m+\bar n-1)!}{R^4(u-v)^{m+n}(\bar{u}-\bar{v})^{\bar m+\bar n}}\ ,
\end{multline}
where $m,\bar m,n,\bar n \geq 1$. Of course, our main task is to understand how this
formula is modified after we have inserted $\Omega$ and integrated over its
insertion point.
In order to spell out the answer, we first define the shorthand
\beq\notag
\calF_{m\bar mn\bar n}\defeq \frac{t_i\otimes t_j\otimes t_k\otimes t_l\left\{([t^l,t^j],[t^k,t^i])(-1)^{|j||k|}+([t^l,t^i],[t^k,t^j])(-1)^{|i|(|k|+|j|)}\right\}m!\bar{m}!n!\bar{n}!}{(z-u)^{m+1}(\bar{z}-\bar{u})^{\bar m+1}(z-v)^{n+1}(\bar{z}-\bar{v})^{\bar n+1}}\ ,
\eeq
which is made out of two Wick contractions and it is designed to provide a useful building block for the four field
correlation function with an insertion of the perturbing field,
\begin{multline}
\vac{\partial^{m}\phi(u)\otimes  \bar{\partial}^{\bar m}\phi(\bar{u})\otimes\partial^{n}\phi(v)\otimes\bar{\partial}^{\bar n}\phi(\bar{v})\,\Omega(z,\bar{z})}_0=
\frac{1}{6R^6}\left(\calF_{m\bar m(n-1)(\bar n-1)}+{}\right.\\\left.{}+\Pi_{24}\cdot\calF_{m(\bar m-1)(n-1)\bar n}+\Pi_{13}\cdot \calF_{(m-1)\bar mn(\bar n-1)}+\Pi_{13}\Pi_{24}\cdot\calF	_{(m-1)(\bar m-1)n\bar n}\right)\ .\label{eq:basic_ertion}
\end{multline}
Here, $\Pi_{ij}$ is the graded permutation operator of the $i$-th and $j$-th
factors of the tensor product. To compute the integral over $z$ of the above
expression, we use the following formula that we derive in app.~\ref{sec:integralidentities},
\begin{multline}
\label{eq:integralid}
\int_{\mathbb{C}_{\epsilon}}\frac{d^2z}{\pi}\frac{a! b! c! d!}{(z-u)^{a+1}(z-v)^{b+1}(\bar{z}-\bar{u})^{c+1}(\bar{z}-\bar{v})^{d+1}}=\\=2\log\left|\frac{u-v}{\epsilon}\right|^2\times \frac{(-1)^{a+c}(a+b)!(c+d)!}{(u-v)^{a+b+1}(\bar{u}-\bar{v})^{c+d+1}}+\text{non-log.}
\end{multline}
With the help of this integral formula and of the Jacobi identity
\beq
\label{eq:JacobiIdentity}
\com{t_i}{\com{t_j}{t_k}}=\com{\com{t_i}{t_j}}{t_k}+(-1)^{|i||j|}\com{t_j}{\com{t_i}{t_k}}\ ,
\eeq
we can evaluate  the term appearing in the second line of eq.~\eqref{eq:sumdecomposition},
\begin{align}\label{eq:intloop1first}
&\int_{\mathbb{C}_{\epsilon}}\frac{d^2z}{\pi}\vac{\partial^{m}\phi(u)\otimes\bar \partial^{\bar{m}}\phi(\bar u)\otimes\partial^{n}\phi(v)\otimes \bar{\partial}^{\bar n}\phi(\bar{v})\,\Omega(z,\bar{z})}_0=2\log\left|\frac{u-v}{\epsilon}\right|^2\times \\
&\;\times (-1)^{|i|(|j|+|\alpha|)}
 [t_\alpha,t_i]\otimes[t^\alpha,t_j]\otimes t^i \otimes t^j \,\frac{(-1)^{n+m}(m+n-1)!(\bar m+\bar n-1)!}{R^6(u-v)^{m+n}(\bar{u}-\bar{v})^{\bar m+\bar n}}+\text{non-log.}\
\notag
\end{align}
The right hand side
of this equation is quite similar to the formula \eqref{eq:basiccorrelator1loop}.
More precisely, the relation is given by
\begin{align}\label{eq:intloop1firstv2}
&\int_{\mathbb{C}_{\epsilon}}\frac{d^2z}{\pi}\vac{\partial^{m}\phi(u)\otimes\bar \partial^{\bar{m}}\phi(\bar u)\otimes\partial^{n}\phi(v)\otimes \bar{\partial}^{\bar n}\phi(\bar{v})\,\Omega(z,\bar{z})}_0=\\
&\;=\frac{2}{R^2} \log\left|\frac{u-v}{\epsilon}\right|^2
(\mathrm{ad}t_\alpha \otimes \mathrm{ad}t^\alpha \otimes \mathbb{1} \otimes \mathbb{1})  \vac{\partial^{m}\phi(u)\otimes  \bar{\partial}^{\bar m}\phi(\bar{u})\otimes\partial^{n}\phi(v)\otimes\bar{\partial}^{\bar n}\phi(\bar{v})}_0
\ , \notag\
\end{align}
where $\mathrm{ad} t_\alpha \defeq [t_\alpha, \cdot\, ]$. We can plug this result back into our basic formula \eqref{eq:sumdecomposition}
to obtain
\begin{multline}\label{eq:op_form_I1}
\int_{\mathbb{C}_{\epsilon}}\frac{d^2z}{\pi}\vac{
\jmath^0_{\tb{m}}(u)\otimes
\bar\jmath^0_{\widebar{\tb{m}}}(\bar{u})\otimes
\jmath^0_{\tb{n}}(v)\otimes \bar\jmath^0_{\widebar{\tb{n}}}(\bar{v})\,\Omega(z,\bar{z})}_0=R^{-2}\log\left|\frac{u-v}{\epsilon}\right|^2\times{}\\{}\times\vac{\left[
\left(\Cas^{\rm D}_{\h}-\Cas^{\rm L}_{\h}\otimes \mathbb{1}_{\rm  R}-\mathbb{1}_{\rm L}\otimes \Cas^{\rm R}_{\h}\right)\cdot
\jmath^0_{\tb{m}}(u)\otimes
\bar\jmath^0_{\widebar{\tb{m}}}(\bar{u})\right]\otimes
\jmath^0_{\tb{n}}(v)\otimes
\bar\jmath^0_{\widebar{\tb{n}}}(\bar{v})
}_0\ .
\end{multline}
In writing this result, we have introduced the Casimir operators $\Cas_{\h}^{\rm L}$, $\Cas_{\h}^{\rm R}$ and $\Cas_{\h}^{\rm D}$ on the representation spaces ${\rm L}\defeq\fm^{\otimes r}$,  ${\rm R}\defeq\fm^{\otimes \bar r}$ and ${\rm D}\defeq{\rm L}\otimes
{\rm R}$, respectively. Note that our expression \eqref{eq:intloop1firstv2} contains
some kind of ``spin-spin interaction'' between the fields with holomorphic and
anti-holomorphic derivatives. This can be expressed in terms of Casimir
elements using the simple identity
\begin{equation}
2\sum_\alpha  t_\alpha \otimes t^\alpha = \sum_\alpha (
t_\alpha \otimes \mathbb{1}+\mathbb{1} \otimes t_\alpha)(
t^\alpha \otimes \mathbb{1}+\mathbb{1} \otimes t^\alpha) - \sum_\alpha t_\alpha t^\alpha \otimes \mathbb{1} -
\sum_\alpha \mathbb{1} \otimes t_\alpha t^\alpha\ .
\end{equation}
Our result \eqref{eq:op_form_I1} may seem to treat the fields at $u$ and $v$ on a different footing.
But  of course the answer is fully symmetric. In fact, using the simple identity
\begin{equation}
(-1)^{|i|(|j|+|\alpha|)}
 [t_\alpha,t_i]\otimes[t^\alpha,t_j]\otimes t^i \otimes t^j = (-1)^{|i|(|j|+|\alpha|)} t_i \otimes t_j\otimes
 [t_\alpha,t^i]\otimes[t^\alpha,t^j]\ ,
\end{equation}
the action of the Casimir operators in eq.~\eqref{eq:op_form_I1} can be moved
from the currents at $u$ to those at $v$.

In order to determine the contribution of $I'_1$ to the 1-loop correction
\eqref{eq:1loopcorrelationfunction} of the 2-point function all that remains
is to take care of the group theoretic factors, and in particular of the
intertwiners ${\sf d}_{\lambda\mu\bar\mu}\otimes {\sf d}_{\xi\eta\bar\eta}$
that appear in the definition of the coset fields \eqref{eq:fieldscosetmodel}.
Their intertwining properties imply that
\begin{multline}
{\sf d}_{\lambda\mu\bar \mu}\, \left(\mathbb{1}_\lambda\otimes\Cas^{\rm D}_{\h}-\mathbb{1}_\lambda\otimes\Cas^{\rm L}_{\h}\otimes \mathbb{1}_{\rm R}-\mathbb{1}_\lambda\otimes\mathbb{1}_{\rm L}\otimes \Cas^{\rm R}_{\h}\right)
={}\\{}=
\left(\Cas^{\lambda}_{\h}\otimes\mathbb{1}_\mu\otimes\mathbb{1}_{\bar\mu}-
\mathbb{1}_\lambda\otimes\Cas^{\mu}_{\h}\otimes \mathbb{1}_{\bar \mu}-\mathbb{1}_\lambda\otimes \mathbb{1}_{\mu}\otimes \Cas^{\bar \mu}_{\h}\right)\, {\sf d}_{\lambda\mu\bar \mu}\ . \label{eq:dCCd}
\end{multline}
We will analyze the implications of this formula in a moment. Before we do
so, let us now turn our attention to the term $I''_1$ in eq.~\eqref{eq:I1I2}
and extract the logarithmic terms. Due to the absence of world-sheet integrals,
the only logarithmic contributions will come from the contraction of $\phi(u,
\bar{u})$ with $\phi(v,\bar{v})$. Concentrating on the relevant factors in eq.\ \eqref{eq:1loopcorrelationfunction} we obtain
\beqa
\label{eq:contributionfromfieldexpansion}
&&\int_{G/H}d\mu(g_0H)\,
\vac{L_\Lambda(\mathrm{Ad}_{g_0}\phi(u,\bar{u}))\calD_{\Lambda\lambda}(g_0)\otimes L_{\Xi}(\mathrm{Ad}_{g_0}\phi(v,\bar{v}))\calD_{\Xi\xi}(g_0)}_0=\nonumber\\&&=
-R^{-2}\log\left|\frac{u-v}{\epsilon}\right|^2\,
\int_{G/H}d\mu(g_0H)\,
L_\Lambda(\mathrm{Ad}_{g_0}t_i)\otimes L_{\Xi}(\mathrm{Ad}_{g_0}t^i)\cdot\calD_{\Lambda\lambda}(g_0)\otimes \calD_{\Xi\xi}(g_0)=\nonumber\\&&=
R^{-2}\log\left|\frac{u-v}{\epsilon}\right|^2\,
\int_{G/H}d\mu(g_0H)\,
L_\Lambda(g_0)L_\Lambda(t_it^i)\calD_{\Lambda\lambda}(e)\otimes \calD_{\Xi\xi}(g_0)
\ ,
\eeqa
where in the second equality we have used the invariance of the scalar product on $L_2(G/H)$ w.r.t.\ the left action of $G$. Next, we use eq.~\eqref{eq:sec_trpr} to
bring this expression to the desired form
\begin{equation}
L_\Lambda(t_it^i) \calD_{\Lambda\lambda}(e) =\left[L_\Lambda(t_it^i+t_\alpha t^\alpha) - R_\lambda(t_\alpha t^\alpha)\right]\calD_{\Lambda\lambda}(e)=\left[\Cas_{\g}^\Lambda - \Cas_{\h}^\lambda\right]\calD_{\Lambda\lambda}(e)\ .
\label{eq:bochner}
\end{equation}
This difference of Casimir elements actually describes the spectrum of the
Bochner Laplacian on the homogeneous vector bundles $B_\lambda$ over
$G/H$, see \cite{Pilch:1984xx}.

Combining the previous results \eqref{eq:op_form_I1} and \eqref{eq:dCCd} with
the expression \eqref{eq:contributionfromfieldexpansion} we conclude that
\begin{multline}
\vac{\Phi_{\boldsymbol{\Lambda}}(u,\bar{u})\otimes \Phi_{\boldsymbol{\Xi}}(v,\bar{v})}_1 =\\=\vac{\frac{1}{R^2}\left(\Cas^{\Lambda}_{\g}-\Cas^{\mu}_{\h}-\Cas^{\bar \mu}_{\h}\right)\cdot \Phi_{\boldsymbol{\Lambda}}(u,\bar{u})\otimes \Phi_{\boldsymbol{\Xi}}(v,\bar{v})}_0\log\left|\frac{\epsilon}{u-v}\right|^2
\label{eq:1loopcorrelationfunctionpart2} + \text{non-log.}
\end{multline}
As in all previous calculations we have ignored all non-logarithmic terms,
which must be removed by proper field redefinitions. The 1-loop anomalous
dimension of the operator \eqref{eq:fieldscosetmodel} can be read off by
comparing eqs.~(\ref{eq:anom_dim_conf}) and
(\ref{eq:1loopcorrelationfunctionpart2}). It is given by
\begin{equation}
\delta \mathbf{h}_{\boldsymbol{\Lambda}}\equiv \delta \mathbf{h}_{\Lambda\lambda\mu\bar{\mu}} = \delta \widebar{\mathbf{h}}_{\Lambda\lambda\mu\bar{\mu}} = \frac{1}{2R^2}\left(\Cas^{\Lambda}_{\g}-\Cas^{\mu}_{\h}-\Cas^{\bar \mu}_{\h}\right)\ .
\label{eq:anom_dim_bos_bulk}
\end{equation}
There are a few comments we would like to make. First, notice that this
formula holds also for $r=0$ or $\bar{r}=0$ and that the dependence on
the label $\lambda$ has dropped out when we added the contributions from
$I'_1$ and $I''_1$. Furthermore, we see that our choice
\eqref{eq:fieldscosetmodel} of a basis in the space of fields in the
coset model diagonalizes the 1-loop anomalous dimensions. Here diagonalization
must be understood in generalized sense, because the representations parametrized
by $\Lambda$, $\mu$, $\bar \mu$ can be indecomposable when we deal with Lie
superalgebras. In such cases the matrices $\Cas^{\lambda}_{\h}$ and
$\Cas^{\mu}_{\h}$, $\Cas^{\mu}_{\h}$ may possess nilpotent terms. These
are to be expected since most conformal field theories on target superspaces
are logarithmic, see e.g. \cite{Schomerus:2005bf} for more explanations.
Finally, let us apply our result \eqref{eq:anom_dim_bos_bulk} to the kinetic
term $(\jmath,\bar{\jmath})$. Since this field is invariant under global $G$ transformations, the representation $\Lambda$ is trivial. The labels $\mu$
and $\bar \mu$, on the other hand, refer to the representation $\mathfrak{m}$
of the multiplets $\jmath,\bar{\jmath}$. Whenever the dual Coxeter number of the
numerator (super)group $G$ vanishes it follows from $[\fm,\fm]\subset \h$ that $\Cas_{\g}^{\g}=2\Cas_{\h}^{\fm}=0$
and hence the anomalous dimension of the kinetic term is zero. This is the
case of conformal coset $\sigma$-models for which the anomalous dimensions are
simply given by eq.\ \eqref{eq:anom_dim_bos_bulk}. Non-conformal models
require a small correction. This is the subject of the next subsection.

\subsubsection{Non-conformal case}

As we discussed in the previous paragraph, the $\sigma$-model will fail to
be conformal at 1-loop if $\Cas_{\h}^{\fm}\neq 0$. If this case there is
an additional contribution to the anomalous dimensions coming from the
wave function renormalization of the field $\phi$ in
eq.~\eqref{eq:wave_func_renorm}
\begin{multline}
\vac{Z_{\Phi_{\boldsymbol{\Lambda}}}\cdot \Phi^{\text{ren}}_{\boldsymbol{\Lambda}}(u,\bar{u})\otimes Z_{\Phi_{\boldsymbol{\Xi}}}\cdot \Phi^{\text{ren}}_{\boldsymbol{\Xi}}(v,\bar{v})}_1 =2(r+\bar{r})\delta Z\vac{\Phi_{\boldsymbol{\Lambda}}(u,\bar{u})\otimes \Phi_{\boldsymbol{\Xi}}(v,\bar{v})}_0{}+\\
+R^{-2}\log\left|\frac{\epsilon}{u-v}\right|^2 \vac{\left(\Cas^{\Lambda}_{\g}-\Cas^{\mu}_{\h}-\Cas^{\bar \mu}_{\h} \right)\cdot \Phi_{\boldsymbol{\Lambda}}(u,\bar{u})\otimes \Phi_{\boldsymbol{\Xi}}(v,\bar{v})}_0\ ,
\label{eq:1loopcorrelationfunctionpart2nonconformal}
\end{multline}
where $\delta Z$ is the 1-loop correction to $Z$ and we have used eq.~\eqref{eq:1loopcorrelationfunctionpart2}. Let us consider the kinetic term   $\form{\jmath}{\bar{\jmath}}$ of the $\sigma$-model for which eq.\
\eqref{eq:wave_func_renorm} reads
\beq
Z_{\form{\jmath}{\bar{\jmath}}}\form{\jmath}{\bar{\jmath}}^{\text{ren}}=Z^2\form{\jmath}{\bar{\jmath}}\ .
\eeq
We want the kinetic term to be exactly marginal and consequently require  $Z_{\form{\jmath}{\bar{\jmath}}}=1$. The 1-loop correction to the 2-point
function of $\form{\jmath}{\bar{\jmath}}^{\text{ren}}$ may be obtained
from eq.~\eqref{eq:1loopcorrelationfunctionpart2nonconformal} by setting
$\Lambda=\Xi=0$ and $\mu=\bar{\mu}=\fm$. Requiring that the cut-off
dependence of $\epsilon$ cancels  and is replaced
by a scale dependence $\mu$, leads to the following formula for the 1-loop wave function renormalization:
\begin{equation}
\delta Z = \frac{\Cas_{\h}^{\fm}}{R^2} \log {\mu} \epsilon+\mathcal{O}(R^{-4})\ .
\label{eq:1loopdeltaZ}
\end{equation}
The prescription \eqref{eq:wave_func_renorm} together with eqs.\ \eqref{eq:1loopcorrelationfunctionpart2nonconformal} and
\eqref{eq:1loopdeltaZ} then implies the following wave function
renormalization for the coset fields \eqref{eq:fieldscosetmodel},
\begin{equation}
Z_{\Phi_{\boldsymbol{\Lambda}}} = 1 + R^{-2}\left[\Cas^{\Lambda}_{\g}-\Cas^{\mu}_{\h}-\Cas^{\bar \mu}_{\h} + (r+\bar r) \Cas_{\h}^{\fm}\right]\log \epsilon {\mu} +\mathcal{O}(R^{-4})\ .
\label{eq:wave_func_ren_comp_fields}
\end{equation}
It is now easy to see that the $\epsilon$ dependence of the 2-point function of  renormalized operators cancels out. Thus, comparing to eq.~\eqref{eq:anom_dim_nonconf} we arrive at the following generalization of eq.~\eqref{eq:anom_dim_bos_bulk} to the non-conformal case
\begin{equation}
\delta \mathbf{h}_{\Lambda\lambda\mu\bar{\mu}} = \delta \widebar{\mathbf{h}}_{\Lambda\lambda\mu\bar{\mu}} = \frac{1}{2R^2}\left[\Cas^{\Lambda}_{\g}-\Cas^{\mu}_{\h}-\Cas^{\bar \mu}_{\h}{} + (r+\bar r) \Cas_{\h}^{\fm}\right]\ .
\label{eq:anom_dim_bos_bulk_nonconf}
\end{equation}
where the scale dependence of the radius $R$ is given by the 1-loop $\beta$-function\footnote{One can use the frame formalism in e.g.\ \cite{Kagan:2005wt} to check that this $\beta$-function agrees with eq.~\eqref{eq:beta_friedan}.} following from eq.~\eqref{eq:def_beta}
\begin{equation}
 {\mu}\frac{d}{d {\mu}}R^2 = 2 \Cas_{\h}^{\fm}\ .
\label{eq:deriv_beta}
\end{equation}
Thus, if $\Cas_{\h}^{\fm}\geq 0$ $(\leq 0)$ then our perturbative calculations will be  valid at  high (low) energy scales $\mu$, where $R^2\to +\infty$ diverges as $|\log \mu|$ and the $\sigma$-model becomes free.

Our result~\eqref{eq:anom_dim_bos_bulk_nonconf} fully agrees  with the formula
obtained by Wegner for the $\mathrm{O}(N)$ vector model \cite{Wegner:1990AD},
i.e.\ the sphere $\sigma$-models. Indeed, if $\Lambda$, $\mu$, $\bar\mu$ are represented by Young diagrams,
then the anomalous dimensions~\eqref{eq:anom_dim_bos_bulk_nonconf} reduce to Wegner's result
\begin{equation}
\delta h_{\Lambda\lambda\mu\bar\mu} =\frac{1}{2R^2}\left[
(N-1)|\Lambda| +(N-2) (r+\bar r- |\mu|-|\bar{\mu}|)+ 2 \xi(\Lambda) - 2\xi(\mu) - 2\xi(\bar\mu)
\right]\ ,
\label{eq:agr_Wegner}
\end{equation}
where for a Young diagram with  rows $Y=(Y_1,Y_2,\dots)$ we have defined $|Y|=\sum_i Y_i$ and $\xi(Y) = \frac{1}{2}\sum_i  Y_i(Y_i-2i+1)$.

\section{Partition functions}\label{sec:pf}

Having computed all 1-loop anomalous dimensions of coset $\sigma$-models it
is tempting to store this information in the partition function. Extending
our earlier discussion of counting functions $\mathcal{Z}_{\Gamma_\lambda}$ for square integrable sections
of compact homogeneous vector bundles, we shall determine a partition function that
counts all fields in the free limit of the compact coset $\sigma$-models, including fields containing an
arbitrary number of derivatives. The expression we find contains some group
theoretic data and it can be deformed very easily to include our results on
1-loop anomalous dimensions. In cases where the group theoretic data is
available, our expressions for the partition function of the free model
can be summed explicitly, as we shall demonstrate in the second subsection
by the example of the quotient $G/H = \mathrm{O}(N)/\mathrm{O}(N-1)=S^{N-1}$. The resulting
expression has a direct geometric meaning which extends to superspheres, see
sec.~\ref{sec:partitionfunctionalternative}.

\subsection{General construction}
\label{sec:gconstr}

In this section we shall  only consider compact $G/H$ coset $\sigma$-models.
The starting point of counting the  states of the coset is given by eq.~\eqref{eq:fieldscosetmodel} for a general coset field that we reproduce here
for the readers convenience
\begin{equation*}
  \Phi_{\boldsymbol{\Lambda}}(z,\bar z) = ({\sf c}_{\lambda\mu\bar\mu}
\mathbb V_{\Lambda\lambda}\otimes
\jmath_{\mu;\tb{m}}\otimes \bar \jmath_{\bar\mu;\widebar{\tb{m}}})(z,\bar z) \  .
\end{equation*}
We have seen in sec.~\ref{sec:1loopbulk} that in the infinite radius limit the dependence of vertex operators $\mathbb V_{\Lambda\lambda}$ on the world-sheet
coordinate drops out, i.e.\ vertex operators are simply square integrable sections, see eq.~\eqref{eq:1loopfields}. At the same time, the currents
$\jmath_{\mu;\tb{m}}$ and $\bar \jmath_{\bar\mu;\widebar{\tb{m}}}$ can be
replaced by derivatives of the fundamental field multiplet as shown in eqs. (\ref{eq:defcurrents2}) and (\ref{eq:defproductcurrents2}). Our proposal for
the space of states of the $\sigma$-model in this limit can be represented in
the following way
\begin{equation}
\mathcal{H}_{G/H} = \left(\,  L_2(G)
\otimes  \mathcal{A}\otimes \bar{\mathcal{A}}\,\right)^{\H-\text{invariants}}
\label{eq:proposal_infr}
\end{equation}
where $L_2(G)$ is viewed as a $G\times H$-bimodule w.r.t.\ the left action of $\G$ and the right action of $\H$, while $\cal A$ and $\bar{\mathcal{A}}$ are the Fock spaces generated by the abelian (in the limit) currents $\jmath$ and $\bar \jmath$, respectively. These carry a left action of $\H$. In case of bosonic groups, eq.~\eqref{eq:proposal_infr} is an obvious consequence of the Peter-Weyl theorem, see the discussion after eq.~\eqref{eq:section}. For supergroups one has to make an assumption about the type of allowed bundles or, more precisely, to specify the
class of indecomposable $H$-representations to which one restricts their fibers.
The factor $L_2(G)$ in eq.~\eqref{eq:proposal_infr} means we restrict to bundles
with projective fibers, see eq.~\eqref{eq:all_bd_pr}.
Clearly, this is a very natural generalization of the bosonic result. Moreover,
as we shall see, the proposal~\eqref{eq:proposal_infr} passes a non-trivial check
at the level of partition functions in the case of supersphere $\sigma$-models.

From our proposal~\eqref{eq:proposal_infr} for the field space of the $\sigma$-model we may read off the infinite radius partition function,
\begin{equation}
\calZ^{\text{free}}_{G/H}(q,\bar q\mid\tb{x}) = \left(\, \calZ_{L_2(G)}(\tb{x}, \tb{y}) \calZ_{\jmath}(q\mid\tb{y})\calZ_{\bar \jmath}(\bar q\mid\tb{y})\,\right)^{\H-\text{invariants}}\ .
\label{eq:pf_gen_pr}
\end{equation}
Here, $\calZ_{L_2(G)}$ is the counting function that was defined in eqs.~(\ref{eq:ch_Glambda}, \ref{eq:pf_all_bd_pr}) and we recall that
$\tb{x}$, $\tb{y}$ are  arbitrary elements of the Cartan tori  of
$\G$ and $\H$, respectively.  The partition function $\calZ_{\jmath}$
is the character of the Fock space generated by $\dim \fm$ bosonic/symplectic
fermionic abelian currents~\eqref{eq:defcurrents2}
\begin{equation}
\calZ_{\jmath}(q\mid\tb{y}) =
\prod_{n=1}^\infty\frac{1}{\mathrm{sdet}[1- R_{\fm}(\tb{y}) q^n]}\ .
\label{eq:pf_ZJ}
\end{equation}
The partition function $\calZ_{\bar \jmath}$ is given by the same formula,
except that the variable $q$ gets replaced by $\bar q$.

The projection on invariants in eq.~\eqref{eq:pf_gen_pr} can be formally
carried out as follows. First, one needs to decompose the partition function $\calZ_{\jmath}$ into the characters $\chi_{W_\mu}$ of the indecomposable
representations $W_\mu$ that are generated by the tensor powers of $\fm$
\begin{equation}
\calZ_{\jmath}(q\mid \tb{y}) = \sum_\mu B_\mu(q) \chi_{W_\mu}(\tb{y}) \ ,
\label{eq:decomp_zj_gen}
\end{equation}
and similarly for $\calZ_{\bar\jmath}$. This expansion defines the branching
functions $B_\mu(q)$ and $B_{\bar \mu}(\bar q)$. The second ingredient we
shall need below is the branching~\eqref{eq:def_mult_sec} of the projective representations of $\G$ into projective representations of $\H$. The latter
determines the decomposition of $\calZ_{L_2(G)}$ into characters of projective representations of $\H$ according to eqs.~(\ref{eq:ch_Glambda}) and
(\ref{eq:pf_all_bd_pr}). Finally, one must evaluate the number
\begin{equation}
\mathcal{N}_{\lambda\mu\bar{\mu}}\defeq \dim\, (P_\lambda\otimes W_\mu\otimes W_{\bar{\mu}})^{H-\text{invariant}}\
\end{equation}
of $\H$-invariants in triple tensor products of $\H$ representations.
In terms of these quantities, we can now rewrite our partition function~\eqref{eq:pf_gen_pr} as follows
\begin{equation}
\calZ^{\text{free}}_{G/H}(q,\bar q\mid\tb{x}) =
\sum_{\Lambda,\lambda,\mu,\bar{\mu}} \chi_{S_\Lambda}(\tb{x}) n_{\lambda\Lambda} \mathcal{N}_{\lambda\mu\bar{\mu}}\, B_\mu(q)B_{\bar{\mu}}(\bar q)\ .
\label{eq:pf_undeformed}
\end{equation}
Since the right hand side keeps track of all the labels that determine the
1-loop anomalous dimensions of bulk fields, we can insert our formula
~\eqref{eq:anom_dim_bos_bulk} into the free partition function to obtain
\begin{equation}
\calZ^{\text{1-loop}}_{G/H}(q,\bar q\mid\tb{x}) =
\sum_{\Lambda,\lambda,\mu,\bar{\mu}} \chi_{S_\Lambda}(\tb{x}) n_{\lambda\Lambda} \mathcal{N}_{\lambda\mu\bar{\mu}}\, B_\mu(q)B_{\bar{\mu}}(\bar q)
(q\bar{q})^{\delta h_{\Lambda\lambda\mu\bar{\mu}}}\ .
\label{eq:pf_deformed}
\end{equation}
Note that all data appearing in this formula, such as the $(G,H)$ branching
numbers $n_{\lambda \Lambda}$, the $H$-fusion rules ${\mathcal N}_{\lambda
\mu\bar \mu}$,
the $G$-characters $\chi_{S_\Lambda}$ and the branching functions $B_\mu$
defined through eqs.\ \eqref{eq:pf_ZJ} and \eqref{eq:decomp_zj_gen} are of
group theoretic nature. Clearly, computing the quantities explicitly is a
highly non-trivial representation theoretic and combinatorial exercise.
In addition, in the supergroup case it requires very good control over
the properties of the indecomposable representations generated by the
tensor powers of $\fm$, which include also the projective representations.
On the other hand, whenever this data is known, our formula
eq.~\eqref{eq:pf_deformed} is not just a formal expression without
practical use. As we shall illustrate in the next subsection, where
we compute explicitly all the group theoretic input of
eq.~\eqref{eq:pf_undeformed} for the $\mathrm{O}(N)$ $\sigma$-models, the summation
over the labels $\Lambda, \lambda, \mu,\bar \mu$ can be carried out to provide
a simple product formula for the free partition function. The final result
possesses a very natural generalization to superspheres. The latter is
derived along a different route in app.~\ref{bulkspectrum}.

\subsection{Sphere and supersphere examples}
\label{sec:ssex}

Let us start by putting the general prescription of the previous section to
work and derive the infinite radius spectrum for $\sigma$-models on spheres.
In this case $\G=\mathrm{SO}(N)$, $\H=\mathrm{SO}(N-1)$ and $\fm$ is the vector representation of $\H$. For sake of simplicity we shall assume that $N$ is even so
that the matrix $R_{\fm}(\tb{y})$ has eigenvalues $(y_A,y^{-1}_A)_{A=1}^{N/2-1}\cup(1)$. The case of $N$ odd is similar and is left as an exercise to the reader. In a slight abuse of notation we shall abbreviate $\tb{y} \equiv R_{\fm}(\tb{y})$. Our first task is to determine the
functions $B_\mu(q)$ which appear when we decompose the function \eqref{eq:pf_ZJ}
into the characters\footnote{For an introduction to the symmetric function $s_{\lambda}$, $sb_{\lambda}$, $sd_{\lambda}$ and their supersymmetric generalizations the reader is referred to \cite{Benkart:2004te} and references therein. } $\chi_{W_\mu} \equiv sb_\mu(\tb{y})$ of irreducible
$\mathrm{SO}(N-1)$ representations. In order to compute the decomposition~\eqref{eq:decomp_zj_gen} we shall extensively use the identity
(4.23) of \cite{Benkart:2004te} (see also \cite{addrem})
\beq
\label{eq:magicsum1}
\sum_{\lambda}sb_{\lambda}(\tb{y})s_{\lambda}(\tb{v})=\frac{\prod_{i\leq j=1}^{\infty}(1-v_iv_j)}{\prod_{j=1}^{\infty}\det (1- \tb{y}v_j)}\ ,
\eeq
where the sum runs over all partitions with at most $N/2-1$ rows, $\tb{v}=(v_1,v_2,\ldots)$ is a vector with possibly infinitely many
complex valued components $v_i$ and $s_{\lambda}$ is the ordinary
Schur function associated to the partition $\lambda$. Using this
identity, the decomposition~\eqref{eq:decomp_zj_gen} reads
\beq
\label{eq:partfuncurretsboundary}
\calZ_\jmath(q\mid \tb{y})=\prod_{i\leq j=1}^{\infty}\frac{1}{(1-q^{i+j})}\sum_{\mu}sb_{\mu}
(\tb{y})s_{\mu}(\tb{q})\ .
\eeq
Here we have defined the vector $\tb{q}=(q, q^2, q^3,\ldots)$ containing
all powers of the variable $q$. Thus, the branching functions for the spheres are
\begin{equation}
B_{\mu}(q)=\prod_{i\leq j=1}^{\infty}(1-q^{i+j})^{-1}s_{\mu}(\tb{q})\ .
\label{eq:br_spheres_expl}
\end{equation}
The partition function for sections on $L_2(\mathrm{SO}(N))$ takes the form
\beq
\label{eq:partfunsections}
\calZ_{L_2(\mathrm{SO}(N))}(\tb{x},\tb{y})=\sum_{\Lambda,\mu}
n_{\lambda\Lambda}sd_{\Lambda}(\tb{x})sb_{\lambda}(\tb{y})\ ,
\eeq
where $\Lambda$ runs over all partitions with at most $N/2$ rows and $sd_{\Lambda}(\tb{x})$ are irreducible $\mathrm{SO}(N)$ characters\footnote{Strictly speaking when $\Lambda$ has exactly $N/2$ rows the symmetric function $sd_{\Lambda}(\tb{x})$ is an $\mathrm{O}(N)$ irreducible character that decomposes into the sum of two $\mathrm{SO}(N)$ irreducible characters.}.
It remains to spell out formulas for the branching numbers $n_{\lambda \Lambda}$
and the fusion rules ${\mathcal N}_{\lambda\mu\bar \mu}$. Both quantities may be
computed from the Littlewood-Richardson coefficients $c^{\lambda}_{\mu\nu}$. For
the branching coefficients $n_{\lambda\Lambda}$ of an $\mathrm{SO}(N)$ \emph{tensor} representation $\Lambda$ into $\mathrm{SO}(N-1)$ representations $\lambda$
the formula
\beq\label{eq:restr_coeff_bn}
n_{\lambda\Lambda}=\sum_{l}c_{\lambda (l)}^{\Lambda}\ ,
\eeq
may be found e.g.\ in  \cite{boerner}. Here, $(l)$ denotes a one row partition
with $l$ boxes. According to the Newell-Littlewood formula, the $\mathrm{SO}(N-1)$ Clebsch-Gordan multiplicities may be obtained as a triple product of Littlewood-Richardson coefficients \cite{King:1971rs},
\beq
\label{eq:NewellLittlewood}
\mathcal{N}_{\lambda\mu\bar{\mu}}=\sum_{\alpha,\beta,\gamma}c^{\lambda}_{\alpha\beta}c^{\mu}_{\beta\gamma}c^{\bar{\mu}}_{\alpha\gamma}\ .
\eeq
Now that we have collected all the representations theoretic data we
can begin to evaluate the general partition function~\eqref{eq:pf_undeformed},
\beqa
\calZ^{\text{free}}_{S^{N-1}}(q,\bar{q}\mid \tb{x})&=&\prod_{i\leq j=1}^{\infty}\frac{1}{|1-q^{i+j}|^2}\sum_{\substack{\Lambda, \lambda,l\\\mu,\bar{\mu},\alpha,\beta,\gamma}}c_{\lambda (l)}^{\Lambda}c^{\lambda}_{\alpha\beta}c^{\mu}_{\beta\gamma}c^{\bar{\mu}}_{\alpha\gamma}sd_{\Lambda}(\tb{x})s_{\mu}(\tb{q})s_{\bar{\mu}}(\bar{\tb{q}})\nonumber\\
&=&\prod_{i\leq j=1}^{\infty}\frac{1}{|1-q^{i+j}|^2}\sum_{\substack{\Lambda, \lambda,l\\\alpha,\beta,\gamma}}c_{\lambda (l)}^{\Lambda}c^{\lambda}_{\alpha\beta}sd_{\Lambda}(\tb{x})s_{\alpha}(\tb{q})s_{\gamma}(\tb{q})s_{\beta}(\bar{\tb{q}})s_{\gamma}(\bar{\tb{q}})\nonumber\\
&=&\prod_{i\leq j=1}^{\infty}\frac{1}{|1-q^{i+j}|^2}\prod_{i,j=1}^{\infty}\frac{1}{(1-q^i\bar{q}^j)}\sum_{\Lambda, \lambda,l}c_{\lambda (l)}^{\Lambda}sd_{\Lambda}(\tb{x})s_{\lambda}(\tb{q},\bar{\tb{q}})\ ,
\label{eq:part_calc1}
\eeqa
where we used the Cauchy identity to evaluate the sum over $\gamma$ and the restriction property $s_\lambda(\tb{u},\tb{v})=\sum_{\alpha,\beta}c^{\lambda}_{\alpha\beta}s_\alpha(\tb{u})s_\beta(\tb{v})$ of the Schur functions to evaluate the sum over $\alpha,\beta$. If we now use the latter identity for a one component vector $\tb{u}=(u)$, then the Schur functions $s_\alpha(\tb{u})$ vanish unless $\alpha=(l)$ is a one row partition in which
case $s_\alpha(\tb{u})=u^l$.  Consequently, we have
\begin{equation}
s_\Lambda(u,\tb{v})=\sum_{\lambda,l}c^{\Lambda}_{\lambda(l)}u^ls_\beta(\tb{v})\ .
\end{equation}
Hence, at the price of introducing a new variable $u$, the sum over $l$ and
$\lambda$ in eq.~\eqref{eq:part_calc1} can also be evaluated
\beqa
\calZ^{\text{free}}_{S^{N-1}}(q,\bar{q}\mid \tb{x})
&=&\lim_{u\to 1}
\prod_{i\leq j=1}^{\infty}\frac{1}{|1-q^{i+j}|^2}\prod_{i,j=1}^{\infty}\frac{1}{(1-q^i\bar{q}^j)}\sum_{\Lambda}sd_{\Lambda}(\tb{x})s_{\Lambda}(u,\tb{q},\bar{\tb{q}})\nonumber\\
{}&=& \left({\lim_{u\rightarrow 1 }}'\frac{1-u^2}{\det(1-\tb{x} u)}\right)\times \prod_{n=1}^{\infty}\frac{|1- q^n|^2}{|\det(1- \tb{x}q^n)|^2}\ ,
\label{eq:part_calc2}
\eeqa
where in the last equality   we applied once again eq.~\eqref{eq:magicsum1} and
$\tb{x}$ is an $\mathrm{SO}(N)$ matrix with eigenvalues $(x_A,x_A^{-1})^{N/2}_{A=1}$. The last limit $u\to 1$, which  naively appears to be zero, must be taken after expanding in powers of $u$. Now, using the formulas
one may find e.g.\ in \cite{Balantekin:1980qy}, one obtains
\begin{equation}
\frac{1-u^2}{\det(1-\tb{x} u)} \bigg\vert_{u=1}= (1-u^2)\sum_{l=0}^\infty u^l h_l(\tb{x}) \bigg\vert_{u=1}= \sum_{l=0}^\infty u^l [h_l(\tb{x})- h_{l-2}(\tb{x})] \bigg\vert_{u=1}= \sum_{l=0}^\infty h'_l(\tb{x})\ ,
\label{eq:part_calc3}
\end{equation}
where $h_l$ are the characters of the $\mathrm{SU}(N)$ symmetric tensors of rank $l$, while $h_l'\defeq h_l- h_{l-2}$ are the characters of $\mathrm{SO}(N)$ \emph{traceless} symmetric tensors.
But these are precisely the representations that appear in the decomposition of $L_2(S^{N-1})$. Hence, we can present the final result for the partition function in the following factorized form
\begin{equation}
\calZ^{\text{free}}_{S^{N-1}}(q,\bar{q}\mid \tb{x}) = \calZ_{L_2(S^{N-1})}(\tb{x})\times \prod_{n=1}^{\infty}\frac{|1- q^n|^2}{|\det(1- \tb{x}q^n)|^2}\ .
\label{eq:final_pf_SN}
\end{equation}
The 1-loop deformation of this partition function is given by eq.~\eqref{eq:pf_deformed} with the group theoretic data listed in eqs.~(\ref{eq:br_spheres_expl}, \ref{eq:restr_coeff_bn}, \ref{eq:NewellLittlewood}) and the anomalous dimensions written in eq.~\eqref{eq:agr_Wegner}.

Eq.~\eqref{eq:final_pf_SN} has a very interesting structure. First, notice that the contribution of zero mode and stringy excitations factorizes.
Such a factorization has been observed before~\cite{Mitev:2008yt}.
Second, notice that the denominator in the second factor corresponds to the partition function of $N$ free bosons in the vector representation of $\mathrm{SO}(N)$.
The numerator, on the other hand, suggests the existence of certain $\mathrm{SO}(N)$ invariant ``null vectors''. These remove all the fluctuations in the embedding space that are transverse to the sphere. With this intuition, we can now very easily generalize the partition function~\eqref{eq:final_pf_SN} to the superspheres $S^{M-1|2N}\defeq \mathrm{OSP}(M|2N)/\mathrm{OSP}(M-1|2N)$
\begin{equation}
\calZ^{\text{free}}_{S^{M-1|2N}}(q,\bar{q}\mid \tb{x}) =\left({\lim_{u\rightarrow 1 }}'\frac{1-u^2}{\sdet(1-\tb{x} u)}\right)\times \prod_{n=1}^{\infty}\frac{|1- q^n|^2}{|\mathrm{sdet}(1- \tb{x}q^n)|^2}\ ,
\label{eq:final_pf_SMN}
\end{equation}
where now $\tb{x}$ is an $\mathrm{OSP}(M|2N)$  matrix.  As in the sphere case, the zero mode contribution is equal to $\calZ_{L_2(S^{M-1|2N})}(\tb{x})$ and can also be extracted from \cite{Candu:2008yw}, where the harmonic analysis on $L_2(S^{M-1|2N})$ was carried out.
The actual calculation of the supersphere partition function is presented in app.~\ref{bulkspectrum}.

\subsection{Alternative derivation}
\label{sec:partitionfunctionalternative}

The goal of this subsection is to present an alternative derivation of the superspheres partition function \eqref{eq:final_pf_SMN} that is not based on our description of the fields in sec.~\ref{subsec:thefields} and that furthermore avoids the subtleties related to the structure of the superalgebra representations.  Specifically, while the details are carefully laid out in app.~\ref{bulkspectrum}, the following comments are only meant to outline the key steps and some underlying ideas.

The starting point is to parametrize the supersphere by an even vector field $X(z,\bar{z})$ taking values in the Euclidean superspace $\mathbb{E}^{M|2N}$ with scalar product $\,\cdot\,$
and subject to the supersphere constraint $X\cdot X=R^2$.
We can implement this constraint by introducing a Lagrange multiplier in the action
\begin{equation}\label{eq:action_LM}
\calS=\int\frac{d^2z}{\pi}\Big[\partial X\cdot \bar{\partial}X+ \lambda(X\cdot X-R^2)\Big]\ .
\end{equation}
The resulting e.o.m.\ $\partial\bar{\partial}X=\lambda X$ together with the constraint $X\cdot X = R^2$ give
\begin{equation}\label{eq:eom_nol}
\partial \bar \partial X = - (\partial X \cdot \bar \partial X) X/R^2\ .
\end{equation}
Next, we trade the field $X(z,\bar{z})$ in favor of its higher order derivatives
\begin{equation}
\label{eq:def_vecs_Xmn}
X_{mn}\defeq \partial^{m}\bar{\partial}^{n}X(z,\bar{z})\big\vert_{z=z_0}\ , \qquad m,n\geq 0\ ,
\end{equation}
at some fixed point $z_0$.
This approach is similar to the concept of ``jet spaces'' used in the BRST quantization of gauge theories, see \cite{Barnich:2000zw} for a review.
Now of course, the vectors $X_{mn}$ are not all algebraically independent. First, one can use the e.o.m.~\eqref{eq:eom_nol} to express the vectors $X_{mn}$ in terms of those in which either $m$ or $n$ is zero. The remaining degrees of freedom $X_{m0}$ and $X_{0n}$ are then subject to polynomial relations  arising from the supersphere constraint
\begin{equation}
 \chi_{mn}\defeq \partial^{m}\bar\partial^{n}\chi(z,\bar{z})\big\vert_{z=z_0}=0\ ,\qquad \chi(z,\bar z)\defeq [X(z,\bar z)\cdot X(z,\bar z) - R^2]\ .
\label{eq:eomc}
\end{equation}
Not all of the constraints $\chi_{mn}$ are independent of each other on-shell.
To understand the dependencies between them  one must relax the condition $\chi(z,\bar z)=0$ and use \emph{only} the e.o.m.
We show in app.~\ref{bulkspectrumnonsusy} that all the constraints $\chi_{mn}$ are linear combinations of those in which either $m$ or $n$ is zero with polynomial coefficients in $X_{mn}$.
Moreover, there are no further relations between  $\chi_{m0}$ and $\chi_{0n}$.
Thus, we have reduced the problem of computing the partition function of the supersphere $\sigma$-model to counting the elements of the polynomial ring $\mathcal{F}$
\emph{freely} generated by $X_{m0}$ and $X_{0n}$
modulo the ideal $\mathcal{I}$ \emph{freely} generated by the constraints $\chi_{m0}$ and $\chi_{0n}$. In other words, we have to count the elements of $\mathcal{F}/\mathcal{I}$.

To proceed further we  use ideas of  BRST cohomology. To every constraint $\chi_{m0}$ and $\chi_{0n}$ we associate a fermionic ghost $c_{m0}$ and $c_{0n}$, respectively,
and then pass to the extended space $\mathcal{C}\defeq \mathcal{F}^*\otimes \mathcal{F}_{\text{gh}}$, where $\mathcal{F}_{\text{gh}}$ is the Grassmann algebra generated by the ghosts and  $\mathcal{F}^*$ is a graded dual of $\mathcal{F}$.
Next, we build
a nilpotent operator $Q:\mathcal{C}\mapsto \mathcal{C}$ whose cohomology we identify  with $(\mathcal{F}/\mathcal{I})^*$.
The partition function of $\mathcal{F}/\mathcal{I}$, which coincides with the partition function of $(\mathcal{F}/\mathcal{I})^*$, is then computed by the following trick.
First, we write down a ``free field'' partition function for $\mathcal{C}$
\begin{equation}
\label{eq:partfunctZtemp}
\calZ_{\calC}(q,\bar{q},u, \{t_{mn}\} \mid \tb{x})=\underbrace{\frac{1}{\sdet(1-u \tb{x} )} \prod_{n=1}^{\infty}\frac{1}{|\sdet(1- u\tb{x}q^n)|^2}}_{\calF^*}\underbrace{(1-t_{00})\prod_{m=1}^{\infty}(1-t_{m0})(1-t_{0m})}_{\calF_{\text{gh}}}\ ,
\end{equation}
where $u$ counts the number  of factors of $X$,  $t_{mn}$ counts the number of ghosts $c_{mn}$ and
$\mathbf{x}$ is an $\OSP{M}{2N}$ matrix keeping track of the transformation properties of the vectors $X_{mn}$.
The trick now is to choose the ghost weights $t_{mn}$ in such a way that the contribution of the preimage and image of $Q$ to the partition function $\mathcal{Z}_{\mathcal{C}}$ cancel against each other.
We argue in app.~\eqref{bulkspectrumnonsusy} that this happens precisely when $t_{mn}=u^2q^m\bar{q}^n$. Plugging these weights into eq.~\eqref{eq:partfunctZtemp} gives exactly the supersphere partition function that we have ``guessed'' in eq.~\eqref{eq:final_pf_SMN}.

\section{World-sheet supersymmetry}
\label{sec:worldsheetsupersymmetry}

In this section, we wish to generalize the results of sec.~\ref{sec:1loopbulk} and \ref{sec:pf} to the case of $\calN=1$ world-sheet supersymmetric $\sigma$-models.
To this end, we first write down the supersymmetric action, give a basis in the space of superfields and then expand these objects at 1-loop.
We then carry out the computation of anomalous dimensions in a manifestly supersymmetric way in complete analogy with the bosonic case, first for conformal and then for non-conformal theories. The dilatation operator turns out to  be formally the same as in the bosonic case.
Then we discuss the general construction of the $\mathcal{N}=1$ partition function of the compact $\sigma$-models to 1-loop and finally illustrate the efficiency of the general approach by the example of the $\mathcal{N}=1$ (super)sphere $\sigma$-models.

\subsection{Background field expansion}

For supersymmetric $\sigma$-models the coordinate fields on the target space get promoted to superfields.
The $\calN=1$ supersymmetrization of the $\sigma$-model Lagrangian~\eqref{eq:lagrangian_bfm}  is obtained in the superspace formalism by sending
\beqa
\label{eq:susysubstitution}
\phi(z,\bar{z})&\mapsto& \phi(Z,\bar{Z})\defeq  \phi(z,\bar{z}) + \theta \psi(z,\bar{z}) + \bar{\theta}\bar{\psi}(z,\bar{z})+\theta\bar{\theta}F(z,\bar{z})\ ,\nonumber\\
\partial&\mapsto& D\defeq \partial_{\theta}-\theta\partial\ ,\nonumber\\
\bar{\partial}&\mapsto& \bar{D}\defeq \bar{\partial}_{\bar{\theta}}-\bar{\theta}\bar{\partial}\ ,
\eeqa
where  $Z=(z,\theta)$, $\bar Z = (\bar z,\bar\theta)$ are superspace coordinates, $\phi(Z,\bar{Z})$ is the $i\fm$ valued coordinate superfield in the coordinate system~\eqref{eq:local_coord} and $F$ is an auxiliary non-dynamical field.
The infinitesimal supersymmetry transformations are then given by
\begin{equation}
\delta_{\epsilon,\bar{\epsilon}} = \epsilon (\partial_\theta+ \theta \partial)+ \bar\epsilon (\partial_{\bar\theta}+ \bar \theta \bar \partial)\ ,
\label{eq:susy_transf}
\end{equation}
where the fermionic derivatives act on the right.
Let us denote the components of the pull-back of the Maurer-Cartan form to the $\mathcal{N}=1$ world-sheet by
\begin{equation}
e^{-i\phi(Z,\bar{Z})} d e^{i \phi(Z,\bar{Z})} = dz J_z  + d\bar z J_{\bar{z}} +d\theta J_\theta  + d\bar \theta J_{\bar\theta} \ ,
\label{eq:components_MC_susy}
\end{equation}
introduce the ``holomorphic'' and ``anti-holomorphic'' current superfields\footnote{These currents play the same role in the $\mathcal{N}=1$ case as the currents \eqref{eq:currdef} played in the $\mathcal{N}=0$ case, hence the same notation.}
\begin{equation}
J\defeq e^{-i \phi(Z,\bar{Z})}D e^{i \phi(Z,\bar{Z})} = J_\theta - \theta J_z\ ,
\qquad
\bar J\defeq e^{-i \phi(Z,\bar{Z})}\bar D e^{i \phi(Z,\bar{Z})} = J_{\bar\theta} - \bar\theta J_{\bar z}
\label{eq:hol_anti_curr_susy}
\end{equation}
and denote their projections on $\fm$ and $\h$ by $\jmath(Z,\bar{Z}) = P J(Z,\bar{Z})$,
$A(Z,\bar{Z}) = P' J(Z,\bar{Z})$ etc.
The supersymmetric action  then reads
\begin{equation}
\mathcal{S} = \frac{R^2}{2}\int \frac{d^2zd\bar\theta d\theta}{\pi} (\jmath ,\bar \jmath)(Z,\bar{Z}) =\frac{R^2}{2}\int\frac{d^2zd\bar\theta d\theta}{\pi}\form{D \phi}{\bar{D}\phi}(Z,\bar{Z})
+\mathcal{S}_{\mathrm{int}}\ ,
\label{eq:action_susy}
\end{equation}
where the perturbing operator at 1-loop is
\begin{equation}
\mathcal{S}_{\mathrm{int}} = \frac{R^2}{2}\int \frac{d^2zd\bar\theta d\theta}{\pi}\Omega(Z,\bar{Z})+\cdots\ ,\qquad \Omega(Z,\bar Z) = \frac{R^2}{6}:([\phi,D\phi],[\phi,\bar D\phi]):(Z,\bar Z)\ .
\label{eq:susy_pert}
\end{equation}
The auxiliary field $F$ does not contribute at 1-loop.
The free propagator takes the standard form
\begin{equation}
\vac{\phi(Z_1,\bar{Z}_1) \otimes \phi(Z_2,\bar{Z}_2)}_0 = -\frac{t_i\otimes t^i}{R^{2}}\log \left|\frac{Z_{12}}{\epsilon}\right|^2\ ,
\label{eq:prop_susy}
\end{equation}
where we have introduced the superinterval  $Z_{12}=z_1-z_2+\theta_1\theta_2$.

The form of a generic superfield can be obtained from eqs.~(\ref{eq:defproductcurrents1}, \ref{eq:fieldscosetmodel}) by making the replacements $\phi(z,\bar{z})\mapsto \phi(Z,\bar{Z})$,
$\jmath(z,\bar{z})\mapsto \jmath(Z,\bar{Z})$ and $\bar{\jmath}(z,\bar{z})\mapsto \bar{\jmath}(Z,\bar{Z})$, where now the covariant superderivatives are defined as $D_A = D + [A,\cdot\,]$\ , $\bar {D}_{\bar A} = \bar D + [\bar A,\cdot\, ]$.
The 1-loop expansion of these fields is given by the simple supersymmetrization of eq.\ \eqref{eq:1loopfields}
\beq
\label{eq:1loopfieldssusy}
\Phi_{\boldsymbol{\Lambda}}(Z,\bar{Z}\mid g_0)=\, :{\sf d}_{\lambda\mu\bar{\mu}}\left[\calD_{\Lambda\lambda}(g_0)+i L_\Lambda(\mathrm{Ad}_{g_0}\phi(Z,\bar{Z}))\calD_{\Lambda\lambda}(g_0)\right]\otimes \jmath^0_{\tb m}(Z)\otimes
\bar \jmath^0_{\widebar{\tb m}}(\bar Z):+\dots\ ,
\eeq
where in the current expansion it sufficient to keep only the dominant terms and the flat part of the covariant superderivatives
\beq
\label{eq:defproductcurrents2_susy}
\jmath^0_{\tb m}(Z)\defeq  \,:\bigotimes_{\rho=1}^r iD^{m_\rho}\phi(Z):\ , \qquad \bar\jmath^0_{\widebar{\tb{m}}}(\bar Z)\defeq \,: \bigotimes_{\rho=1}^{\bar r}i\bar D^{\bar{m}_\rho}\phi(\bar{Z}):\ .
\eeq
We now have all the ingredients to start the calculation of 1-loop anomalous dimensions.

\subsection{1-loop dilatation operator}

Let us first assume that the $\mathcal{N}=1$ $\sigma$-model is conformal.
We then proceed as in the bosonic case~\eqref{eq:1loopcorrelationfunction} by splitting the  1-loop logarithmic correction to the 2-point function of two arbitrary operators $\vac{\Phi_\Lambda(U,\bar U)\otimes \Phi_\Xi(V,\bar{V})}$ into two parts ---
one coming from the expansion of vertex operators, which we call $\vac{\cdot }''_1$, and the other one coming from the insertion of the interaction, which we call $\vac{\cdot}'_1$.
The first one is computed exactly as in eqs.~(\ref{eq:contributionfromfieldexpansion}, \ref{eq:bochner}), but with the new propagator~\eqref{eq:prop_susy}, and gives
\begin{multline}
\vac{\Phi_{\boldsymbol{\Lambda}}(U,\bar{U})\otimes \Phi_{\boldsymbol{\Xi}}(V,\bar{V})}''_1 =\\=\vac{\left[\frac{1}{R^2}\left(\Cas^{\Lambda}_{\g}-\Cas^{\lambda}_{\h}\right)\cdot \Phi_{\boldsymbol{\Lambda}}(U,\bar{U})\right]\otimes \Phi_{\boldsymbol{\Xi}}(V,\bar{V})}_0\log\left|\frac{\epsilon}{Z_{uv}}\right|^2+\text{non-log.}\ ,
\label{eq:1loopcorrelationfunctionpart2_susy}
\end{multline}
where we have introduced the notation $Z_{uv} = u-v + \theta_u \theta_v$ and $\bar{Z}_{uv} = \bar u-\bar v + \bar \theta_u \bar \theta_v$.

To compute the second contribution, the following formula\footnote{Here $\floor{x}\defeq \max \{n\in \mathbb{Z} \mid n\leq x\}$. } is very  useful
\begin{equation}
D_1^m D_2^n \log Z_{12} = - (-1)^{ \floor{n/2}}\floor{(m+n-1)/2 }!Z_{12}^{-(m+n)/2}\ ,
\label{eq:useful_susy_form}
\end{equation}
where we have introduced the standard convention for the half-integer power of the superinterval $Z^{-n-1/2}_{12} \defeq  (\theta_1-\theta_2)Z^{-n-1}_{12}$.
With these conventions, the combinatorics of Wick contractions in the supersymmetric case becomes identical  to the bosonic case, up to some grading signs due to the odd parity of $D$, $\bar{D}$.
Eq.~\eqref{eq:basiccorrelator1loop} is now replaced by
\begin{multline}
\label{eq:basiccorrelator1loop_susy}
\vac{D^{m}\phi(U)\otimes  \bar{D}^{\bar m}\phi(\bar{U})\otimes D^{n}\phi(V)\otimes\bar{D}^{\bar n}\phi(\bar{V})}_0=\\=
(-1)^{|i||j|}t_i \otimes t_j\otimes t^i\otimes t^j\,\frac{(-1)^{\bar m n +\floor{n/2}+\floor{\bar n/2}}\floor{(m+n-1)/2}!\floor{(\bar m+\bar n-1)/2}!}{R^4Z_{uv}^{(m+n)/2}\bar{Z}_{uv}^{(\bar m+\bar n)/2}}\ .
\end{multline}
The basic Wick contraction~\eqref{eq:basic_ertion} gets decorated by signs
\beqa
\label{eq:basiccorr_susy}
&&\vac{D^{m}\phi(U)\otimes  \bar{D}^{\bar m}\phi(\bar{U})\otimes D^{n}\phi(V)\otimes\bar{D}^{\bar n}\phi(\bar{V})\,\Omega(Z,\bar{Z})}_0\nonumber\\&&\hspace{1cm}=\frac{1}{6R^6}\Big[(-1)^{\bar m}\calF_{m\bar m(n-1)(\bar n-1)}+(-1)^{\bar m + n+\bar n}\Pi_{24}\cdot\calF_{m(\bar m-1)(n-1)\bar n}+\nonumber\\&&\hspace{1.5cm}+(-1)^{n+1}\Pi_{13}\cdot \calF_{(m-1)\bar mn(\bar n-1)}+
(-1)^{\bar n}\Pi_{13}\Pi_{24}\cdot\calF	_{(m-1)(\bar m-1)n\bar n}\Big]\ ,
\eeqa
where now
\begin{multline}
\calF_{m\bar mn\bar n}\defeq t_i\otimes t_j\otimes t_k\otimes t_l\left\{([t^l,t^j],[t^k,t^i])(-1)^{|j||k|}+([t^l,t^i],[t^k,t^j])(-1)^{|i|(|k|+|j|)}\right\}\times\\ \times
\frac{\floor{m/2}!\floor{\bar{m}/2}!\floor{n/2}!\floor{\bar{n}/2}!}{Z_{uz}^{(m+1)/2}\bar Z_{uz}^{(\bar m+1)/2} Z_{vz}^{(n+1)/2}\bar Z_{vz}^{(\bar n+1)/2}}\ .
\end{multline}
The integration over the insertion point can be carried out with the help of the following generalization of eq.~\eqref{eq:integralid}:
\begin{multline}
\int_{\mathbb{C}^{1|1}_{\epsilon}} \frac{d^2z d\bar\theta d\theta}{\pi}\frac{\floor{a/2}!\floor{b/2}!\floor{c/2}!\floor{d/2}!}{Z_{uz}^{(a+1)/2}\bar{Z}_{uz}^{(b+1)/2}Z_{vz}^{(c+1)/2}\bar{Z}_{vz}^{(d+1)/2}} =\\= 2\log \left|\frac{Z_{uv}}{\epsilon}\right|^2\times \frac{(-1)^{bc+\floor{(c+1)/2}+\floor{(d+1)/2}}\floor{\frac{a+c}{2}}!\floor{\frac{b+d}{2}}!}{Z_{uv}^{(a+c+1)/2}\bar Z_{uv}^{(b+d+1)/2}}+ \text{non log.}\ ,
\label{eq:basic_integral_susy}
\end{multline}
where $\mathbb{C}^{1|1}_\epsilon\defeq \{ (z,\theta)\mid\epsilon\leq|z-u|,  |z-v|\}$. Applying this formula and the Jacobi identity  \eqref{eq:JacobiIdentity}   we get
\begin{align}\label{eq:intloop1first_susy}
&\int_{\mathbb{C}^{1|1}_{\epsilon}}\frac{d^2zd\bar\theta d\theta}{\pi}\vac{D^{m}\phi(U)\otimes\bar D^{\bar{m}}\phi(\bar U)\otimes D^{n}\phi(V)\otimes \bar{D}^{\bar n}\phi(\bar{V})\,\Omega(Z,\bar{Z})}_0=\frac{2}{R^2}\log\left|\frac{Z_{uv}}{\epsilon}\right|^2\times  \notag\\
&\times (-1)^{|i|(|j|+|\alpha|)}
 [t_\alpha,t_i]\otimes[t^\alpha,t_j]\otimes t^i \otimes t^j \,\frac{(-1)^{\bar m n +\floor{n/2}+\floor{\bar n/2}}\floor{\frac{m+n-1}{2}}!\floor{\frac{\bar m+\bar n-1}{2}}!}{R^4 Z_{uv}^{(m+n)/2}\bar Z_{uv}^{(\bar m+\bar n)/2}}\ .
\end{align}
Comparing to the free correlator \eqref{eq:basiccorrelator1loop_susy} and repeating the Wick combinatorics of the bosonic case we arrive at the desired correction
\begin{multline}
\vac{\Phi_{\boldsymbol{\Lambda}}(U,\bar{U})\otimes \Phi_{\boldsymbol{\Xi}}(V,\bar{V})}'_1 =\\=\vac{\left[\frac{\Cas^{\lambda}_{\h}-\Cas^{\mu}_{\h}-\Cas^{\bar \mu}_{\h}}{R^2}\cdot \Phi_{\boldsymbol{\Lambda}}(U,\bar{U})\right]\otimes \Phi_{\boldsymbol{\Xi}}(V,\bar{V})}_0\log\left|\frac{\epsilon}{Z_{uv}}\right|^2\ .
\label{eq:1loopcorrelationfunctionpart2_susy_2}
\end{multline}
The two contributions~\eqref{eq:1loopcorrelationfunctionpart2_susy} and \eqref{eq:1loopcorrelationfunctionpart2_susy_2} are exactly analogous to the bosonic case.
Summing them up we get the same formal expression~\eqref{eq:anom_dim_bos_bulk} for the  (operatorial)  anomalous dimension.

For non-conformal theories, just as in the bosonic case, there is an additional contribution to the anomalous dimensions coming from the renormalization of the coordinates fields themselves.
The renormalization procedure is identical to the bosonic case and leads again to the same form~\eqref{eq:anom_dim_bos_bulk_nonconf} for the anomalous dimensions.

\subsection{Partition functions}

This subsection is the $\mathcal{N}=1$ supersymmetric generalization of sec.~\ref{sec:pf}. First we discuss the general construction and then focus on the example of $\mathcal{N}=1$ superspheres.

\subsubsection{General construction}

The general approach of sec.~\ref{sec:gconstr} to computing the partition functions of compact $\sigma$-models in the infinite radius limit generalizes straightforwardly to the $\mathcal{N}=1$ case.
The coset fields have the same structure as  before
\begin{equation*}
  \Phi_{\boldsymbol{\Lambda}}(Z,\bar Z) = {\sf c}_{\lambda\mu\bar\mu}\, \cdot
\mathcal{D}_{\Lambda\lambda}[\phi(Z,\bar Z)]\otimes
\jmath_{\mu;\tb{m}}(Z,\bar{Z})\otimes \bar \jmath_{\bar\mu;\widebar{\tb{m}}}(Z,\bar{Z})\ .
\end{equation*}
Hence, in the large radius limit the space of states $s\mathcal{H}_{G/H}$ of the $\mathcal{N}=1$ theory reduces to
\begin{equation*}
s\mathcal{H}_{G/H} = \left(\,  L_2(G)
\otimes  s\mathcal{A}\otimes s\bar{\mathcal{A}}\,\right)^{\H-\text{invariants}}\ ,
\end{equation*}
where now  $s\cal A$  is the Fock space generated by the abelian (in the limit) currents $PJ_z$ together with their free fermion (in the limit) superpartners $PJ_\theta$, and $s\bar{\mathcal{A}}$ is constructed similarly out of the barred quantities.
We can now repeat the discussion of sec.~\ref{sec:gconstr} to arrive at the partition function of the $\mathcal{N}=1$ coset (computed with an insertion of $(-1)^{F+\bar F}$)
\begin{equation}
s\calZ^{\text{1-loop}}_{G/H}(q,\bar q\mid\tb{x}) =
\sum_{\Lambda,\lambda,\mu,\bar{\mu}} \chi_{S_\Lambda}(\tb{x}) n_{\lambda\Lambda} \mathcal{N}_{\lambda\mu\bar{\mu}}\, sB_\mu(q)sB_{\bar{\mu}}(\bar q)
(q\bar{q})^{\delta h_{\Lambda\lambda\mu\bar{\mu}}}\ .
\label{eq:pf_deformed_super}
\end{equation}
simply by including the contribution of  fermions in the partition function for the supercurrents
\begin{equation}
s\calZ_{\jmath}(q\mid\tb{y}) =
\prod_{n=1}^\infty\frac{\mathrm{sdet}[1- R_{\fm}(\tb{y}) q^{n-\frac{1}{2}}]}{\mathrm{sdet}[1- R_{\fm}(\tb{y}) q^n]}
=\sum_\mu sB_\mu(q) \chi_{W_\mu}(\tb{y})\ .
\label{eq:pf_sZJ}
\end{equation}

\subsubsection{$\mathcal{N}=1$ superspheres}

The $\mathcal{N}=1$ generalization of the large radius $\mathrm{O}(N)/\mathrm{O}(N-1)$ sphere partition function calculation of sec.~\ref{sec:ssex} is based on the supersymmetric version of the combinatorial identity~\eqref{eq:magicsum1}
\begin{equation}
\frac{\prod_j \det (1-\mathbf{y}w_j)}{\prod_j \det (1-\mathbf{y}v_j)} \prod_{i\leq j}(1-v_i v_j)\prod_{i< j}(1-w_i w_j) = \prod_{i,j}(1-v_i w_j)\sum_\mu sb_\mu(\mathbf{y}) s_\mu(\mathbf{v}|\mathbf{w})\ ,
\label{eq:supermagic}
\end{equation}
which we prove in app.~\ref{sec:comb_id}.
Here $\mathbf{y}$ is an $\SO{N-1}$ matrix (with $N$ even), $\mathbf{v}=(v_1,v_2,\dots)$ and $\mathbf{w}=(w_1,w_2,\dots)$ are possibly infinite vectors, and $s_\mu(\mathbf{v}|\mathbf{w})$ are the supersymmetric Schur functions,\footnote{If $\mathbf{v}$ has $k$ components and $\mathbf{w}$ has $l$ components, then the symmetric functions $s_\mu(\mathbf{v}|\mathbf{w})$ can be interpreted as the characters of $\mathrm{gl}(k|l)$ covariant tensors.} see eq.~\eqref{eq:def_sschur} for definitions. We only need to know that they  satisfy the properties  $s_\lambda s_\mu = \sum_\nu c_{\lambda\mu}^\nu s_\nu$ and $\sum_{\mu\nu}c^{\lambda}_{\mu\nu}s_{\mu}(v|w)s_{\nu}(v^{\prime}|w^{\prime})=s_{\lambda}(v,v^{\prime}|w,w^{\prime})$, similarly to the usual Schur functions.
Setting  the vectors $\mathbf{v}=(q,q^2,q^3,\dots)$ and $\mathbf{w}=(q^{\frac{1}{2}},q^\frac{3}{2},q^\frac{5}{2},\dots)$ in eq.~\eqref{eq:supermagic} we get  for the branching functions in the decomposition~\eqref{eq:pf_sZJ}
\begin{equation}
sB_\mu(q) = \frac{\prod^\infty_{i,j=1}(1-q^{i+j-\frac{1}{2}})}{\prod_{i\leq j=1}^\infty(1-q^{i+j})\prod_{i< j=1}^\infty(1-q^{i+j-1})}\times s_\mu(\mathbf{v}|\mathbf{w})\ .
\label{eq:s_branch}
\end{equation}
Next, one can repeat the arguments of sec.~\ref{sec:ssex} word by word until one arrives at
\begin{equation}
s\mathcal{Z}^{\text{free}}_{S^{N-1}}(q,\bar{q}\mid \mathbf{x}) = \mathcal{Z}_{L_2(S^{N-1})}(\mathbf{x})\times \prod_{n=1}^\infty\frac{|1-q^n|^2}{|1-q^{n-\frac{1}{2}}|}\frac{|\mathrm{det}(1-\mathbf{x}q^{n-\frac{1}{2}})|^2}{|\mathrm{det}(1-\mathbf{x}q^{n})|^2}\ ,
\label{eq:pf_n1_spheres_gen}
\end{equation}
where  $\mathbf{x}$ is an $\mathrm{SO}(N)$ matrix.
The natural guess for superspheres is then
\begin{equation}
s\mathcal{Z}^{\text{free}}_{S^{M-1|2N}}(q,\bar{q}\mid \mathbf{x}) = \mathcal{Z}_{L_2(S^{M-1|2N})}(\mathbf{x})\times \prod_{n=1}^\infty\frac{|1-q^n|^2}{|1-q^{n-\frac{1}{2}}|^2}\frac{|\sdet(1-\mathbf{x}q^{n-\frac{1}{2}})|^2}{|\sdet(1-\mathbf{x}q^{n})|^2}\ ,
\label{eq:pf_n1_spheres_gen_super}
\end{equation}
where now $\mathbf{x}$ is an $\mathrm{OSP}(M|2N)$ matrix.
Ultimately, the partition function~\eqref{eq:pf_n1_spheres_gen_super} is justified by the honest direct calculation in app.~\ref{bulkspectrum}.

\section{Discussions and Outlook}
\label{sec:discussionsoutlook}

In this article, we computed the anomalous dimensions in both compact and non-compact $\sigma$-models
with symmetric target spaces at 1-loop in perturbation theory around
the infinite radius point. We performed these computations in cases
without and with $\calN=1$ world-sheet supersymmetry. To this end, we
introduced a special basis \eqref{eq:fieldscosetmodel}
for the bulk fields, in which the anomalous dimensions are given by
eq.~\eqref{eq:anom_dim_bos_bulk_nonconf}. For conformal $\sigma$-models,
the expression reduces to eq.\ \eqref{eq:anom_dim_bos_bulk}.
In the presence of world-sheet supersymmetry, the formulas
\eqref{eq:anom_dim_bos_bulk_nonconf} and \eqref{eq:anom_dim_bos_bulk}
continue to hold. The only difference with the bosonic case is that
world-sheet fermions and their derivatives can contribute to the
representations $\mu$ and $\bar \mu$ of the denominator group as
well as to the integers $r$ and $\bar r$ that appear in
non-conformal models.

In addition to computing anomalous dimensions for all bulk fields,
we were able
to assemble this information in a formula for the 1-loop partition
function of the compact $\sigma$-models on  symmetric superspaces. This
partition function accounts for all states that possess finite
conformal dimension in the infinite radius limit. Such states may
be thought of as momentum states, in contrast to winding states
which, in case they appear, acquire infinite conformal weight
as we send the $\sigma$-model coupling to zero. It would be very
interesting to investigate under which conditions such 1-loop
partition functions are modular invariant. In case they are not,
one should be able to extract information on the winding sector.
Recently, Douglas and Gao have initiated a similar analysis for
$\sigma$-models on Calabi-Yau spaces \cite{Gao:2013mn}. Through an
interesting link with trace formulas they argued that the modular
invariance of the 1-loop partition function requires the
presence of winding states. Conformal $\sigma$-models on symmetric
superspaces may cast a new light on the issue.

There are many other interesting open problems that should be
addressed. The most relevant ones concern applications to $\sigma$-models for strings in AdS backgrounds which do not satisfy our
basic assumptions. Most importantly, while being coset spaces
their targets cease to be symmetric. Instead, the denominator
subgroup $H$ is kept fixed by an automorphism of order four.
In addition, the relevant models possess fermionic Wess-Zumino
terms, the ``metric'' in the action is degenerate (in the Green-Schwarz formalism) in the fermionic sector due to $\kappa$-symmetry and their target spaces are necessarily non-compact. It
would certainly be important to include all these features
into our analysis. Our construction of fields goes through for
more general classes of coset spaces and it seems likely that
the corresponding basis continues to diagonalize the 1-loop
dilatation operator. On the other hand, the interaction term
takes a more complicated form so that several contributions
need to be added in evaluating the perturbed 2-point functions.
Wess-Zumino terms bring in additional contributions to the
interaction. Let us note in passing that, independently of
applications to the AdS/CFT correspondence, it would be
desirable to incorporate $\theta$-terms, i.e.\ $B$-fields
that are described by a closed 2-form without global 1-form potential. These appear e.g.\ in $\mathbb{CP}^{n-1|n}$
models and they have an effect on the bulk spectrum. To take them into account, one should generalize the background field method to classical backgrounds with non-trivial instanton charge. Finally,
non-compactness of the quotient space modifies our discussion
in case the denominator group $H$ is non-compact because of
issues with normalizability. When $H$ is not compact, harmonic
analysis on $G$, which has been our main input in the
construction and enumeration of tachyon vertex operators,
ceases to be a suitable starting point for the construction
of normalizable states on $G/H$. All these are technical
complications one should be able to overcome with limited
efforts.

Another aspect that deserves further study is the duality
between sigma and WZW models. One such duality between the
$\sigma$-model on a supersphere $S^{2n+1|2n}$ and the
$\OSP{2n+2}{2n}$ WZW model at level $k=1$ was described and
analyzed in \cite{Candu:2008yw,Mitev:2008yt} through the study
of conformal weights for boundary fields. More recently, we
determined the conformal weights for a subsector of bulk
fields in perturbed WZW models to all-loop order
\cite{Candu:2012xc}. These expressions should be compared
with the results we have described above. In particular,
it would be very interesting to identify the corresponding
subsector in the $\sigma$-model. Since our expressions for
anomalous dimensions in deformed WZW models only involve
the quadratic Casimir element of the group $G$ of global
symmetries, it is tempting to think that non-derivative
fields of the $\sigma$-model might play an important role in
the correspondence. Indeed, the quadratic Casimir element
of the denominator group $H$ only enters the expression
\eqref{eq:anom_dim_bos_bulk_nonconf} through the
transformation law of the currents $\jmath$.

Further directions include the computation of anomalous
dimensions to higher orders in perturbation theory and an
analysis of boundary fields for symmetry preserving boundary
conditions. For boundary fields, the above formulas are
expected to simplify significantly and higher or even all-loop
computations of anomalous dimensions become more feasible,
see also \cite{Quella:2007sg, Mitev:2008yt,Candu:2009ep} for some existing studies in this direction. We plan to come back to some of these issues in future research.

\section*{Acknowledgments}
The authors wish to thank Alessandra Cagnazzo, Tigran Kalaydzhyan,
Hubert Saleur, Thomas Spencer and Christoph Sieg for comments and
interesting discussions. They are especially indebted to Yuri
Aisaka for some inspiring early discussions that helped shaping
the construction of fields. The research leading to these results
has received funding from the People Programme (Marie Curie Actions)
of the European Union's Seventh Framework Programme FP7/2007-2013/
under REA Grant Agreement No 317089 (GATIS).

\appendix

\section{Free supersphere  spectrum}
\label{bulkspectrum}

\index{bulkspec1}

In this appendix, we shall present a method for the computation of bulk spectra of $S^{M-1|2N}$ supersphere $\sigma$-models in the infinite radius limit. In \cite{Mitev:2008yt},  the problem of computing the boundary spectra was solved combinatorially in the particular case of  $S^{3|2}$ superspheres, from which a general formula was guessed.
Here, we shall present a different counting method based on cohomology that applies equally well to all superspheres $S^{M-1|2N}$ both in the bulk and boundary cases with or without world-sheet supersymmetry.
For definiteness, we shall  consider only bulk partition functions.

In principle, the method presented in this section applies to all NLSM which admit a presentation as constrained free theories.

\subsection{$\calN=0$ case}
\label{bulkspectrumnonsusy}

In sec.~\ref{sec:partitionfunctionalternative},
we claimed that the space of states of the supersphere $\sigma$-model can be identified with  $\mathcal{H}\defeq\mathcal{F}/\mathcal{I}$, where  $\mathcal{F}$ is the polynomial ring
freely\footnote{The generators of a polynomial ring are free if they do not satisfy any non-trivial polynomial relations.} generated by the components of the vectors
(see eq.~\eqref{eq:def_vecs_Xmn})
\begin{equation}\label{eq:trading}
X\equiv X_{00}\ ,\quad X_m\equiv X_{m0} \ ,\quad \bar X_m \equiv X_{0m} \ ,\qquad (m\geq 1)
\end{equation}
modulo the ideal $\mathcal{I}$ freely generated by the constraints (see eq.~\eqref{eq:eomc})
\begin{equation}\label{eq:constr_app}
\chi\equiv \chi_{00}\ ,\quad \chi_m\equiv \chi_{m0}\ ,\quad \bar \chi_m \equiv \chi_{0m}\ ,\qquad (m\geq 1)\ .
\end{equation}
The only point that remains to be proven is that all the constraints $\chi_{mn}$ can be expressed in terms of those in which either $m$ or $n$ is zero, i.e.\ $\chi_{mn}\in\mathcal{I}$. This can be proved as follows.
Relaxing the constraint $\chi(z,\bar z)=0$ and using \emph{only} the e.o.m.~\eqref{eq:eom_nol} we get
\begin{equation}
\partial \bar\partial \chi(z,\bar z) = 2(\partial X\cdot \bar\partial X + X\cdot \partial\bar\partial X)(z,\bar z) = -2 R^{-2}(\partial X\cdot \bar \partial X)(z,\bar z)\chi (z,\bar z)\ .
\label{eq:basic_eq}
\end{equation}
Next, let us denote by $\mathcal{J}_{mn}$ the space of fields generated by
\begin{equation}
\mathcal{J}_{mn}\defeq \langle \chi(z,\bar z), \partial \chi(z,\bar z),\dots, \partial^m \chi(z,\bar z), \bar\partial \chi(z,\bar z),\dots, \bar \partial^n \chi(z,\bar z)\rangle \ .
\end{equation}
Then, using the basic equation \eqref{eq:basic_eq} one derives $\partial \mathcal{J}_{mn}\subset \mathcal{J}_{(m+1)n}$ and $\bar \partial \mathcal{J}_{mn}\subset \mathcal{J}_{m(n+1)}$.
Hence, if  $m,n>0$ we have $\partial^m\bar\partial^n \chi(z,\bar z)\subset \partial^{m-1}\bar\partial^{n-1}\mathcal{J}_{00}\subset \mathcal{J}_{(m-1)(n-1)}$.
Evaluating this last relation at $z=z_0$ we get the desired result  that $\chi_{mn}$ belongs to the ideal of $\cal F$ generated by $\chi$, $\chi_1,\dots,\chi_{m-1}$, $\bar \chi_1$, \dots, $\bar\chi_{n-1}$.

Let us now define and compute the partition function of $\mathcal{H}$.
For this purpose we shall need the following gradings on $\mathcal{F}$\footnote{All fermionic derivatives act on the right.}
\beq
 \label{eq:gradings1}
\begin{split}
N&\defeq X^a\frac{\partial}{\partial X^a}+ \frac{\partial}{\partial R}\ ,\quad N_{m}\defeq\sum_{a=1}^{M+2N} X^a_{m}\frac{\partial}{\partial X^a_{m}}\ , \quad
\bar N_{m}\defeq\sum_{a=1}^{M+2N} \bar X^a_{m}\frac{\partial}{\partial \bar X^a_{m}}\\
D_a &\defeq  X^a\frac{\partial }{\partial X^a}+\sum_{m=1}^{\infty}\left( X^a_{m} \frac{\partial}{\partial X^a_{m}}+\bar X^a_{m} \frac{\partial}{\partial \bar X^a_{m}}\right)\ ,
\end{split}
\eeq
where $R$ is the supersphere radius, i.e.\ $X\cdot X = R^2$, and the components of every vector $V=e_a V^a$ are taken w.r.t.\ a complex basis $(e_a)$ of the Euclidean superspace $\mathbb{E}^{M|2N}$ diagonalizing the action of some fixed Cartan subalgebra of $\osp{M}{2N}$.
Thus, $N$ counts the number of $X^{a}$ or $R$ factors, $N_m$ ($\bar N_m$) the number of $X^{a}_m$ ($\bar X^{a}_m$) factors and $D_a$ the number of $X^a$, $X^a_{m}$ or $\bar X^a_{m}$ factors.
We also introduce the total number and energy operators
\beq\label{eq:gradings2}
 N_t\defeq N + \sum_{m=1}^\infty (N_m +\bar N_m) \ ,\quad E\defeq\sum_{m=1}^{\infty}mN_{m}\ ,\quad  \bar{E}\defeq\sum_{m=1}^{\infty}m\bar N_{m}\ .
\eeq
Notice that the generators  of the ideal $\mathcal{I}$ in eq.~\eqref{eq:constr_app} have only definite $N_t$, $E$ and $\bar{E}$-degrees ---
the $(N_t,E,\bar{E})$ degree of $\chi$ is $(2,0,0)$, of $\chi_m$ is $(2,m,0)$ and of $\bar\chi_m$ is $(2,0,m)$.
This means that the gradings defined by $N_t$, $E$ and $\bar E$ descend to the quotient space $\mathcal{H}$.
Hence, we have the following decomposition
\begin{equation}
\calF = \bigoplus_{r,e,\bar{e}=0}^\infty \calF^{r,e,\bar{e}}\ , \qquad  \calH=\bigoplus_{r,e,\bar{e}=0}^\infty \calH^{r,e,\bar{e}}\ ,
\end{equation}
where  $\calF^{r,e,\bar{e}}$ and $\calH^{r,e,\bar{e}}$ are finite dimensional subspaces of degree $(r,e,\bar{e})$ w.r.t.\ $(N_t,E,\bar{E})$ and we have $\calH^{r,e,\bar{e}}=\calF^{r,e,\bar{e}}/\calI^{r,e,\bar{e}}$ with $\calI^{r,e,\bar{e}}=\calF^{r,e,\bar{e}}\cap \calI$.
The supersphere partition function is then defined by
\begin{multline}\label{eq:pf_def}
 \mathcal{Z}^{\text{free}}_{S^{M-1|2N}}(q,\bar{q}|u,\textbf{x})\equiv \calZ(q,\bar{q}|u,\textbf{x})\defeq \str_{\mathcal{H}} u^{N_t} q^E \bar q^{\bar E} \prod_{a=1}^{M+2N} x_a^{D_a} = \\=
\sum_{r,e,\bar{e}=0}^\infty u^r q^e \bar{q}^{\bar{e}}\left[\str_{\calF^{r,e,\bar{e}}} \left(\prod_{a=1}^{M+2N} x_a^{D_a}\right)- \str_{\calI^{r,e,\bar{e}}} \left(\prod_{a=1}^{M+2N} x_a^{D_a}\right)\right]\ .
\end{multline}
where $\mathbf{x}$ is an element of the Cartan subgroup of $\text{OSP}(M|2N)$ with eigenvalues $\mathbf{x}\cdot e_a = x_a e_a$.

In order to compute the  partition function we find it convenient to pass to the dual spaces $\calF^*$ and $\calH^*$, which are defined in the graded sense
\begin{equation}\label{eq:grad_dual}
\calF^* \defeq \bigoplus_{r,e,\bar{e}=0}^\infty (\calF^{r,e,\bar{e}})^*\ ,\qquad \calH^* \defeq \bigoplus_{r,e,\bar{e}=0}^\infty (\calH^{r,e,\bar{e}})^* \ .
\end{equation}
We shall assign to $(\calF^{r,e,\bar{e}})^*$ the same $N_t$, $E$  and $\bar{E}$ degrees as $\calF^{r,e,\bar{e}}$.
Then, due to the reality of the $\text{OSP}(M|2N)$-representation in which the generators $X$, $X_m$, $\bar X_m$ transform, there is a natural isomorphism $\calH\simeq \calH^*$
of  $\text{OSP}(M|2N)$ modules. Hence we can compute the partition function as
\begin{equation}\label{eq:pf_general}
\calZ(q,\bar{q}|u,\mathbf{x})=  \str_{\calH^*}u^{N_t} q^E \bar{q}^{\bar{E}}  \prod_{a=1}^{M+2N} x_a^{D_a}\ .
\end{equation}
The advantage of passing to the dual spaces is that we can characterize  $\mathcal{H}^*$ as the subspace of $\calF^* $ vanishing on  $\calI$.
Now, recalling that  $\calI$ is generated by $\chi$, $\chi_m$ and $\bar\chi_m$ we get
\begin{equation}\label{eq:def_ker}
\calH^* = \Ker  q^o\cap \bigcap_{m=1}^\infty (\Ker q^o_m\cap \Ker \bar q^o_m)\ ,
\end{equation}
where
\begin{equation}\label{eq:ev_map}
(q^o \cdot t)(p) \defeq  t(\chi p)\ ,\quad    (q^o_m \cdot t)(p) \defeq  t(\chi_m p)\ ,\quad
(\bar q^o_m \cdot t)(p) \defeq  t(\bar \chi_m p)\ .
\end{equation}
It is easy to see  that the maps $q^o$, $q^o_m$, $\bar q^o_m$ are $\text{OSP}(M|2N)$-invariant, change the degrees as
\begin{equation}
(\calF^{r,e,\bar{e}})^* \stackrel{q^o}{\longrightarrow}  (\calF^{r-2,e,\bar{e}})^*\ ,\quad (\calF^{r,e,\bar{e}})^* \stackrel{q^o_m}{\longrightarrow}  (\calF^{r-2,e-m,\bar{e}})^*\ ,
\quad
(\calF^{r,e,\bar{e}})^* \stackrel{\bar q^o_m}{\longrightarrow}  (\calF^{r-2,e,\bar{e}-m})^*\ ,
\label{eq:change_grad_ev}
\end{equation}
 and, most importantly,  are \emph{surjective}, which  follows  from the injectivity of the dual maps, i.e.\ multiplication by $\chi$, $\chi_m$, $\bar\chi_m$.

\begin{figure}[h]
\centerline{\includegraphics[trim = 0cm 2cm 0cm 0cm, clip, height=6cm]{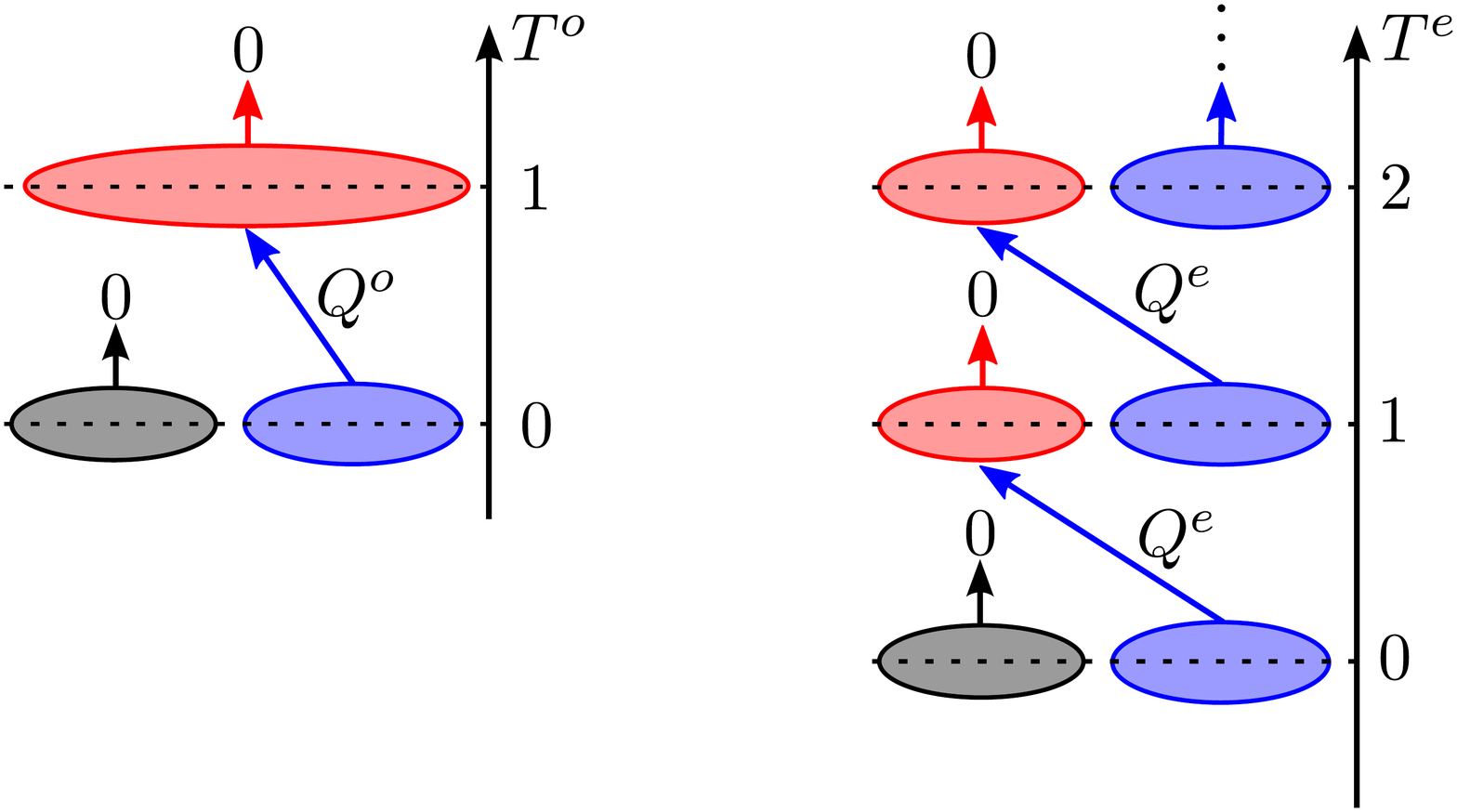}}
\caption{The action of the  nilpotent operators $Q^o$ and $Q^e$ associated to a fermionic and bosonic ghost, respectively. The preimage of is drawn blue, the image in red.
\label{fig:cohom}
}
\end{figure}

This fact can be used to reduce the calculation of the partition function~\eqref{eq:pf_general} to a cohomological problem.
Thus, we associate to every constraint in eq.~\eqref{eq:constr_app} a fermionic ghost: to $\chi$ the ghost $c$,  to $\chi_m$ the ghost $c_m$, to $\bar \chi_m$ the ghost $\bar c_m$.
Then we pass to the extended space
\begin{equation}\label{eq:extended_sp}
\calC = \calF^{*}\otimes \calF_{\text{gh}}\ ,
\end{equation}
where $\calF_{\text{gh}}$ is the Grassmann algebra generated by the ghosts.
With respect to the previous  gradings $N_t$, $E$, $\bar E$ and the total ghost number operator
\begin{equation}
T^o_t = T^o + \sum_{m=1}^\infty (T^o_m + \bar T^o_m)\  ,
\label{eq:total_gh_nb}
\end{equation}
where
\beq
\label{eq:ghost_nb}
T^o \defeq c\frac{\partial}{\partial c}\ ,\quad  T^o_m \defeq c_m\frac{\partial}{\partial c_m}\ ,\quad \bar T^o_m \defeq \bar c_m\frac{\partial}{\partial \bar c_m}\ ,
\eeq
the extended space~\eqref{eq:extended_sp} decomposes as
\begin{equation}
\calC= \bigoplus_{r,e,\bar{e},p=0}^\infty\calC^{r,e,\bar{e}}_p
\end{equation}
where $p$ is the total number of ghosts.
We now turn this space into a complex with respect to the nilpotent map $Q^o_t: \calC_p \mapsto \calC_{p+1}$
defined as follows
\begin{equation}\label{eq:def_qs}
Q^o_t \defeq Q^o +\sum_{m=1}^\infty (Q^o_m+\bar Q^o_m) \ ,\qquad Q^o  \defeq q^o \otimes c, \quad  Q^o_m  \defeq q^o_m \otimes c_m, \quad \bar Q^o_m  \defeq \bar q^o_m \otimes \bar c_m, \quad
\end{equation}
It is easy to check that  $Q^o$, $Q^o_{m}$ and $\bar Q^o_m$  anticommute with each other. Therefore, the cohomology of $Q^o_t$ lies in the combined cohomology of the former.
Moreover, the surjectivity of $q^o$ implies that the cohomology of $Q^o$ lies exclusively at $T^o$-ghost number 0 and coincides exactly with $\Ker q^o$.
A similar statement holds for the cohomology of $Q^o_m$ and $\bar Q^o_m$.
This is represented in fig.~\ref{fig:cohom}.
Thus, we arrive at the desired cohomological reformulation of eq.~\eqref{eq:def_ker}
\begin{equation}\label{eq:beauty}
\Coh^{Q^o_t}_p \defeq \frac{\Ker Q^o_t: \calC_p \mapsto \calC_{p+1}}{\Img Q^o_t: \calC_{p-1} \mapsto \calC_{p}} = \delta_{p0}\calH^*\ .
\end{equation}
Eq.~\eqref{eq:beauty} can be used very efficiently to compute the partition function~\eqref{eq:pf_general} by means of the following trick.
First,  we compute the ``free field'' partition function over  $\calC$
\begin{multline}
\label{eq:partitionfunctionextended}
\calZ_{\calC}\defeq \str_\calC\left[ u^{N_t} q^{E}\bar{q}^{\bar{E}}  t^{T^o} \prod_{m=1}^{\infty} (t_{m}^{T^o_{m}}\bar t_m^{\bar T^o_m}) \prod_{a}x_a^{D_a}\right]=
\frac{1-t}{\sdet(1-\mathbf{x}u)}\prod_{m=1}^{\infty}\frac{(1-t_m)(1-\bar{t}_m)}{|\sdet(1-\mathbf{x} u q^m)|^2}\ .
\end{multline}
Next, we consider the decomposition of  this partition function w.r.t.\ the action of $Q^o$
\beq \label{eq:idea}
\calZ_{\calC}= \calZ_{\mathsf{H}}+
\calZ_{\mathsf{Im}}+ \calZ_{\mathsf{Im}^{-1}}=
\calZ_{\mathsf{H}} + \calZ_{\mathsf{Im}}(1-t^{-1} u^{2})\ ,
\eeq
where $\calZ_{\mathsf{H}}$,
$\calZ_{\mathsf{Im}}$ and $\calZ_{\mathsf{Im}^{-1}}$ respectively denote the contribution
of states in the cohomology, image and preimage of $Q^o$.
We could express $Z_{\mathsf{Im}^{-1}}$ in terms of  $Z_{\mathsf{Im}}$ since
\begin{align}
&\com{T}{Q^o}=Q^o\ ,& &\com{N}{Q^o}=-2Q^o\ ,& \nonumber &\com{E}{Q^o}=0\ ,& &\com{\bar{E}}{Q^o}=0\ ,&
\label{eq:rel_im_preim}
\end{align}
which follows from eqs.~(\ref{eq:change_grad_ev}, \ref{eq:ghost_nb}).
The weight  $t$ keeps track of the ghost variable $c$ and is clearly unphysical.
We can set $t=u^{2} $ and obtain from eq.~\eqref{eq:idea}
\beq
\label{reductiontrick}
\calZ_{\calC}\big\vert_{t=u^{2} } = Z_{\Coh}\big\vert_{t=u^{2} }=\calZ_{\Coh^{Q^o}_0}\ .
\eeq
The last equality is valid since the cohomology of $Q^o$ is localized exclusively at $T^o$-ghost number 0, hence it is independent of $t$.
A similar argument applied to $Q^o_{m}$, $\bar Q^o_m$ gives
\beq\label{eq:restr_coh_qmbqm}
\mathcal{Z}_{\mathcal{C}}\big\vert_{t_{m}=u^2q^{m}}=\calZ_{\Coh^{Q^o_m}_0}\ , \qquad \mathcal{Z}_{\mathcal{C}}\big\vert_{\bar t_{m}=u^2\bar q^{m}}=\calZ_{\Coh^{\bar Q^o_m}_0}\ .
\eeq
The partition function in the combined cohomology is obtained by specializing the partition function~\eqref{eq:partitionfunctionextended} to the ghost weights appearing in eqs.~(\ref{reductiontrick}, \ref{eq:restr_coh_qmbqm}).
This leads to the final result
\begin{align}
\label{partitionfunctioninfvolwithoutsusy}
\mathcal{Z}^{\text{free}}_{S^{M-1|2N}}(q,\bar{q}|u,\textbf{x})=\frac{1-u^2}{\sdet(1- \mathbf{x} u )}\prod_{n=1}^{\infty}\frac{|1-u^2 q^n|^2}{|\sdet(1- \mathbf{x} u q^n)|^2}\ .
\end{align}

\subsection{$\calN=1$ case}
\label{bulkspectrumsusy}

We shall apply the new method for the computation of free spectra that we have explained in detail in the previous section to the case of $\calN=1$ world-sheet supersymmetric $S^{M-1|2N}$ superspheres.
According to~\cite{Witten:1977xn}, we  can replace in the action~\eqref{eq:action_LM} all fields by superfields
\begin{equation}
\begin{split}
X(z,\bar z)&\mapsto X(Z,\bar{Z})= X(z,\bar{z})+\theta Y(z,\bar{z})+\bar{\theta} \bar{Y}(z,\bar{z})+\theta\bar{\theta}F(z,\bar{z})\ , \\
\lambda(z,\bar{z})&\mapsto \lambda(Z,\bar{Z})=G(z,\bar{z})+\theta\sigma(z,\bar{z})+\bar{\theta}\bar{\sigma}(z,\bar{z})+\theta\bar{\theta}\lambda(z,\bar{z})
\label{eq:all_fields_n1}
\end{split}
\end{equation}
and then integrate over the fermionic superspace variables  to get the supersymmetric action
\begin{multline}
\label{eq:susyaction}
\calS=\int\frac{d^2z}{\pi}\left[\partial X\cdot \bar{\partial} X+ Y\cdot \bar{\partial}Y+ \bar{Y}\cdot \partial\bar{Y}+ F\cdot F +\lambda(X\cdot X-R^2)+{}\right.\\\left.{}+2\bar \sigma Y\cdot X-2\sigma \bar{Y}\cdot X
+2G( F\cdot X-Y\cdot \bar{Y})\right]\ .
\end{multline}
Next, we use the e.o.m.\ for the entire system~\eqref{eq:all_fields_n1} to eliminate the auxiliary field $F$ and the Lagrange multipliers $G$, $\sigma$, $\bar\sigma$ and $\lambda$. The output consists of the constraints
\beq
\label{eq:susyconstraintssusy}
\chi\defeq X\cdot X-R^2=0\ , \quad \psi\defeq X\cdot Y=0\ ,\quad \bar\psi \defeq X\cdot \bar Y=0\ ,
\eeq
together with the  e.o.m.\ for  $X$, $Y$ and $\bar Y$
\beq
\label{eq:eom_n1_ssph}
\begin{split}
\partial \bar \partial X &= - [ (\partial X\cdot \bar \partial X)X+(\partial X \cdot \bar Y)\bar Y+(\bar \partial X\cdot Y)Y]/R^2\ ,\\
\bar \partial  Y &= -[(Y\cdot \bar\partial X)X+(Y\cdot \bar Y)\bar Y]/R^2\ ,\quad \partial \bar Y =-[(\bar Y\cdot \partial X)X+(\bar Y\cdot Y)Y]/R^2\ .
\end{split}
\eeq

We shall now repeat the same steps as in the $\mathcal{N}=0$ case.
First, we trade the world-sheet dependence of the fields $X$, $Y$, $\bar Y$ for the  values of their higher order derivatives~\eqref{eq:trading} and
\begin{equation}
Y_m \defeq \partial^m Y(z,\bar z)\vert_{z=z_0} \ ,\quad \bar Y_m \defeq \bar \partial^m \bar Y(z,\bar z)\vert_{z=z_0} \ ,\qquad (m\geq 0)
\label{eq:trade_f}
\end{equation}
at some fixed point $z=z_0$. On-shell, all other  derivatives can be expressed in terms of these ones using the e.o.m.~\eqref{eq:eom_n1_ssph}.
Next,  we introduce a polynomial ring $s\mathcal{F}$ freely generated by the components of $X$, $X_m$, $\bar X_m$, $Y_m$ and $\bar Y_m$ and an ideal $s\mathcal{I}\subset s\mathcal{F}$
freely generated by the constraints~\eqref{eq:constr_app} and
\begin{equation}
\psi_m \defeq \partial^m \psi(z,\bar z)\vert_{z=z_0}\ ,\quad \bar \psi_m \defeq \bar \partial^m  \bar\psi(z,\bar z)\vert_{z=z_0}\ ,\qquad (m\geq 0)\ .
\label{eq:gen_constr_ferm_ind}
\end{equation}
One can now identify the space of states of the $\mathcal{N}=1$ supersphere $\sigma$-model with the quotient $s\mathcal{H}\defeq s\calF/s\calI$, because  on-shell the most general derivatives of the constraints~\eqref{eq:susyconstraintssusy}
are  linear combinations of the free generators~(\ref{eq:constr_app}, \ref{eq:gen_constr_ferm_ind}) of $s\mathcal{I}$ with coefficients in $s\calF$. The last part of the claim is proved as in the $\mathcal{N}=0$ case
using the basic equations
\begin{equation}
\begin{split}
\partial\bar\partial \chi(z,\bar z)&= - \frac{2}{R^2}\left[(\partial X\cdot \bar\partial  X )\chi+(\partial X\cdot \bar Y) \bar \psi +(\bar\partial  X\cdot Y)\psi \right](z,\bar z)\ ,\\
\bar \partial \psi  (z,\bar z) &= -\frac{1}{R^2} \left[( Y\cdot \bar \partial X)\chi +( Y\cdot \bar Y)\bar \psi \right](z,\bar z)\ ,\\
\partial \bar \psi  (z,\bar z) &= -\frac{1}{R^2} \left[(\bar Y\cdot \partial X)\chi +(\bar Y\cdot  Y)\psi \right](z,\bar z)\  .
\end{split}
\end{equation}
Then we pass to the dual space $s\calH^*$, which one can characterize as
\begin{equation}
s\calH^*\simeq \Ker  q^o\cap \bigcap_{m=1}^\infty (\Ker q^o_m\cap \Ker \bar q^o_m)\cap \bigcap_{m=0}^\infty (\Ker q^e_m\cap \Ker \bar q^e_m)\ ,
\label{eq:coh_ref_sh}
\end{equation}
where $q^e_m$ and $q^e_m$ are  the maps dual to the multiplication by $\psi_m$ and $\bar \psi_m$, respectively.
Notice that the maps $q^e_m$ and $q^e_m$ are exact, because their duals are exact, i.e.\ the only elements of $s\calF$ in the kernel of the multiplication by the odd constraints $\psi_m$ ($\bar \psi_m$) are those which are proportional to $\psi_m$ ($\bar \psi_m$). This observation, together with eq.~\eqref{eq:coh_ref_sh} can be used to give a cohomological reformulation for $s\calH$ just as in the $\mathcal{N}=0$ case.

Thus, besides the fermionic ghosts $c$, $c_m$, $\bar c_m$ introduced in the previous section, we associate to all fermionic constraints $\psi_m$ and $\bar\psi_m$ the
corresponding bosonic ghosts $\gamma_m$ and $\bar\gamma_m$, and pass to the extended space $s\calC\defeq s\calF^*\otimes s\calF_{\text{gh}}$, where  $s\calF_{\text{gh}}$ is the polynomial ring generated by all ghosts.
We then turn $s\calC$ into a complex w.r.t.\ the nilpotent map
\begin{equation}
Q_t \defeq Q^o_t + \sum_{m=0}^\infty (Q^e_m+\bar Q^e_m)\ ,\qquad Q^e_m\defeq q^e_m\otimes\gamma_m\ ,\qquad \bar Q^e_m\defeq \bar q^e_m\otimes\bar \gamma_m\ .
\label{eq:}
\end{equation}
Once again, the cohomology of $Q_t$ lies in the combined cohomology of its elementary  summands, because the latter anticommute with each other. Moreover, the exactness   of $q^e_m$ ($\bar q^e_m$) implies that the cohomology of $Q^e_m$ ($\bar Q^e_m$) also lies at $\gamma_m$-ghost number 0, see fig.~\ref{fig:cohom}. Thus, on the one hand, we conclude that the cohomology of $Q_t$ lies exclusively at total ghost number 0, while on the other hand eq.~\eqref{eq:coh_ref_sh} (see also fig.~\ref{fig:cohom}) implies that it coincides with $s\calH^*$.

One can now use the same trick as in the previous section to derive the partition function of $s\calH\simeq s\calH^*$ from the ``free field'' partition function of $s\calC$
\begin{equation}\label{eq:pf_sc_n1}
s\calZ_{s\calC}= \frac{1-t}{\sdet(1-\mathbf{x}u)}\prod_{m=1}^{\infty}\frac{(1-t_m)(1-\bar{t}_m)}{|\sdet(1-\mathbf{x} u q^m)|^2}\times \prod_{m=0}^\infty\frac{|\sdet(1-\mathbf{x}u q^{m+\frac{1}{2}})|^2}{(1-\tau_m)(1-\bar \tau_m)}\ ,
\end{equation}
where every bosonic ghost $\gamma_m$ ($\bar{\gamma}_m$) is weighted by a factor of $\tau_m$ ($\bar \tau_m$).
Arguing as before, one concludes that by setting the ghost weights to the special values~(\ref{reductiontrick}, \ref{eq:restr_coh_qmbqm}) and
\begin{equation}
\tau_{m}=u^2q^{m+\frac{1}{2}}\ ,\quad \bar{\tau}_{m}=u^2\bar{q}^{m+\frac{1}{2}}\ ,
\end{equation}
only the elements in the  cohomology of $Q_t$ contribute.
Inserting these values in eq.~\eqref{eq:pf_sc_n1} we obtain the final result
\begin{equation}
\label{eq:bulkpartitionfunctionsusyappendix}
s\mathcal{Z}^{\text{free}}_{S^{M-1|2N}}(q,\bar{q}\mid \mathbf{x})=\frac{1-u^2}{\sdet(1- \mathbf{x} u )}\prod_{n=1}^{\infty}\left|\frac{1-u^2 q^n}{1-u^2q^{n-\frac{1}{2}}}\frac{\sdet(1-\mathbf{x}uq^{n-\frac{1}{2}})}{\sdet(1- \mathbf{x} u q^n)}\right|^2\ .
\end{equation}

\section{An integral identity}
\label{sec:integralidentities}

For the reader's convenience, we have assembled in this appendix the derivation of the integral formula  \eqref{eq:integralid}. For this, we use Stokes' theorem and arrive at
\beqa
\int_{\mathbb{C}_{\epsilon}}\frac{d^2z}{\pi}\frac{1}{(z-x)(z-y)(\bar{z}-\bar{x})(\bar{z}-\bar{y})}&=&\frac{1}{\bar{x}-\bar{y}}\int_{\mathbb{C}_{\epsilon}}\frac{d^2z}{\pi}\bar{\partial}_z\frac{\log\left|\frac{z-x}{z-y}\right|^2}{(z-x)(z-y)}\nonumber\\ &=&-\frac{1}{\bar{x}-\bar{y}}\oint_{\partial\mathbb{C}_{\epsilon}}\frac{dz}{2\pi i}\frac{\log\left|\frac{z-x}{z-y}\right|^2}{(z-x)(z-y)}\ ,
\eeqa
where in the last line the contour integral has two counterclockwise oriented pieces -- the first around $x$ and the second around $y$ -- both of which give the same contribution. Performing this integrals is easy and leads to
\beq
\int_{\mathbb{C}_{\epsilon}}\frac{d^2z}{\pi}\frac{1}{(z-x)(z-y)(\bar{z}-\bar{x})(\bar{z}-\bar{y})}=\frac{2\log\left|\frac{x-y}{\epsilon}\right|^2}{|x-y|^2}+\calO(\epsilon)\ .
\eeq
By taking now the appropriate number of derivatives in $x$, $y$, $\bar{x}$ or $\bar{y}$ on both sides of the above equation, we recover eq.~\eqref{eq:integralid}.

\section{A combinatorial identity}\label{sec:comb_id}
\label{app:combinatorial}

In this appendix we shall prove the  combinatorial identity~\eqref{eq:supermagic}, namely
\begin{equation*}
\frac{\prod_j \det (1-\mathbf{y}w_j)}{\prod_j \det (1-\mathbf{y}v_j)} \prod_{i\leq j}(1-v_i v_j)\prod_{i< j}(1-w_i w_j) = \prod_{i,j}(1-v_i w_j)\sum_\mu sb_\mu(\mathbf{y}) s_\mu(\mathbf{v}|\mathbf{w})\ ,
\end{equation*}
where $\mathbf{y}$ is an $\SO{2n+1}$ matrix, $\mathbf{v}=(v_1,v_2,\dots)$ and $\mathbf{w}=(w_1,w_2,\dots)$ are possibly infinite vectors and $s_\mu(\mathbf{v}|\mathbf{w})$ are the supersymmetric Schur functions defined by the expansion
\begin{equation}
\prod_{i,j}\frac{(1-v_i w'_j)(1-v'_i w_j)}{(1-v_i v'_j)(1-w_i w'_j)} = \sum_\mu s_\mu(\mathbf{v}|\mathbf{w}) s_\mu(\mathbf{v}'|\mathbf{w}')\ .
\label{eq:def_sschur}
\end{equation}
They  satisfy the defining  property  $s_\lambda s_\mu = \sum_\nu c_{\lambda\mu}^\nu s_\nu$ of the Schur functions and can be expressed in terms of the ``bosonic'' Schur functions as
\begin{equation}
s_\lambda(\mathbf{v}|\mathbf{w})= \sum_{\mu,\nu} c^\lambda_{\mu\nu} s_\mu(\mathbf{v})s_{\nu^t}(-\mathbf{w})\ ,
\label{eq:restr_formula}
\end{equation}
see eqs.~(A.7) and (A.10) of \cite{Candu:2012jq}.
We shall also need the identities \cite{MacDonald}
\begin{equation}
\prod_{i\leq j}(1-v_iv_j) = \sum_{\rho\in R} (-1)^{\frac{|\rho|}{2}}s_\rho(\mathbf{v})\ ,\qquad \prod_{i< j}(1-w_iw_j) = \sum_{\rho\in R} (-1)^{\frac{|\rho|}{2}}s_{\rho^t}(\mathbf{w})\ ,
\label{eq:mcdonald_ref}
\end{equation}
where $|\rho|$ is the number of boxes in $\rho$ and $R$ is the set of partitions for which every $i$-th row is one box longer then the respective $i$-th column. In particular, all $\rho\in R$ have an even number of boxes.
Notice that together eqs.~(\ref{eq:magicsum1}, \ref{eq:mcdonald_ref})  imply the well known relation between the $\SO{2n+1}$ and $\mathrm{SU}(2n+1)$ characters, see e.g.\ \cite{addrem},
\begin{equation}
sb_\lambda(\mathbf{y}) = \sum_{\mu}\sum_{\rho\in R} (-1)^{\frac{|\rho|}{2}}c^\lambda_{\rho\mu} s_\mu(\mathbf{y})\ .
\label{eq:rel_char_so_su}
\end{equation}
Furthermore, using eq.~\eqref{eq:mcdonald_ref}, the ``dual'' Cauchy identity
\begin{equation}\label{eq:magicsum2}
\prod_{j}\det (1+\mathbf{y} w_j) = \sum_\lambda s_\lambda(\mathbf{y})s_{\lambda^t}(\mathbf{w})\ ,
\end{equation}
and then the relation \eqref{eq:rel_char_so_su} one can prove\footnote{We remind of the properties $c_{\mu\nu}^{\lambda}=c_{\mu^t\nu^t}^{\lambda^t}$ and $c_{\mu\nu}^{\lambda}c_{\rho\sigma}^{\mu}=c_{\mu\rho}^{\lambda}c_{\sigma\nu}^{\mu}=c_{\mu\sigma}^{\lambda}c_{\nu\rho}^{\mu}$. } the  ``dual'' of eq.~\eqref{eq:magicsum1}
\begin{equation}
\prod_j \det(1+\mathbf{y}w_j)\prod_{i<j}(1-w_iw_j) = \sum_\lambda sb_\lambda(\mathbf{y})s_{\lambda^t}(\mathbf{w})\ .
\label{eq:dual_cauchy}
\end{equation}
The identity~\eqref{eq:supermagic} can now be proved as follows
\begin{align*}
&\frac{\prod_j \det (1-\mathbf{y}w_j)}{\prod_j \det (1-\mathbf{y}v_j)} \prod_{i\leq j}(1-v_i v_j)\prod_{i< j}(1-w_i w_j) =
\sum_{\lambda,\mu}sb_\lambda(\mathbf{y}) sb_\mu(\mathbf{y}) s_\lambda(\mathbf{v})s_{\mu^t}(-\mathbf{w})  =\\
&=\mathop{\sum_{\lambda,\mu,\nu}}_{\alpha,\beta,\gamma} c^{\lambda}_{\alpha\beta}c^\mu_{\beta\gamma}c^\nu_{\gamma\alpha}sb_\nu(\mathbf{y})s_\lambda(\mathbf{v})s_{\mu^t}(-\mathbf{w})
=\mathop{\sum_{\nu,\alpha}}_{\beta,\gamma}
c^\nu_{\gamma\alpha} sb_\nu(\mathbf{y}) s_\alpha(\mathbf{v})s_\beta(\mathbf{v})s_{\beta^t}(-\mathbf{w})s_{\gamma^t}(-\mathbf{w}) =\\
&= \prod_{i,j}(1-v_i w_j)\sum_{\alpha,\gamma,\nu}c^\nu_{\gamma\alpha}sb_\nu (\mathbf{y})s_\alpha(\mathbf{v})s_{\gamma^t}(-\mathbf{w})= \prod_{i,j}(1-v_i w_j) \sum_\nu sb_\nu(\mathbf{y})s_\nu(\mathbf{v}|\mathbf{w})\ ,
\end{align*}
were the first equality follows from eqs.~(\ref{eq:magicsum1}, \ref{eq:magicsum2}), in the second equality we have used the Newell-Littlewood formula~\eqref{eq:NewellLittlewood}, in the fourth equality the dual Cauchy identity~\eqref{eq:dual_cauchy}  and in the last equality the restriction formula~\eqref{eq:restr_formula} for the supersymmetric
Schur functions.


\end{document}